\documentclass[12pt]{iopart}
\usepackage{iopams}
\usepackage{epsfig}
\input{epsf}
\begin{document}
\title{Quantum cosmology and late-time singularities}
\author{A Yu Kamenshchik}
\address{
Dipartimento di Fisica e Astronomia and INFN, Sezione di Bologna, Via Irnerio 46, 40126 Bologna, Italy\\
L.D. Landau Institute for Theoretical Physics of the Russian Academy of Sciences, Kosygin str.~2, 119334 Moscow, Russia}
\eads{\mailto{kamenshchik@bo.infn.it}}
\begin{abstract}
The development of dark energy models has stimulated interest to cosmological singularities, which differ from the traditional 
Big Bang and Big Crunch singularities. We review a broad class of phenomena connected with soft cosmological singularities in 
 classical and quantum cosmology. We discuss the classification of singularities from 
the geometrical point of view and from the point of view of the behaviour of finite size objects, crossing such singularities.
We discuss in some detail quantum and classical cosmology of models based on  perfect fluids (anti-Chaplygin gas and anti-Chaplygin gas plus dust), of models based on the Born-Infeld-type fields and of the model of a scalar field with a potential inversely proportional to the field itself. We dwell also on the phenomenon of the phantom divide line crossing in the scalar field models with cusped potentials. Then we discuss the Friedmann equations modified by quantum corrections to the effective action of the models under considerations and the influence of such modification on the nature and the existence of soft singularities. We review also quantum cosmology of models, where the initial quantum state of the universe is presented by the density matrix (mixed state). Finally, we discuss the exotic singularities arising in the brane-world cosmological models. 
\end{abstract}
\submitto{CQG}
\maketitle
\section{Introduction}
The problem of cosmological singularities has been attracting the attention
of theoreticians working in gravity and cosmology since the early fifties 
\cite{Land,Misn-Torn,Hawk-Ell}. In the sixties general theorems about the
conditions for the appearance of singularities were proven \cite{Hawk,Pen}
and the oscillatory regime of approaching the singularity \cite{BKL}, called also 
``Mixmaster universe'' \cite{Misner} was discovered. Basically, until the end of
nineties almost all discussions about singularities were devoted to the Big
Bang and Big Crunch singularities, which are characterized by a vanishing
cosmological radius.

However, kinematical investigations of Friedmann cosmologies have raised the
possibility of sudden future singularity occurrence \cite{sudden1}, 
characterized by a diverging $\ddot{a}$ whereas both the scale
factor $a$ and $\dot{a}$ are finite. Then, the Hubble parameter $H=\dot{a}%
/a~ $and the energy density $\rho $ are also finite, while the first
derivative of the Hubble parameter and the pressure $p$ diverge. 
Until recent years, however, the sudden future singularities attracted rather a limited interest 
of researchers. The situation has changed drastically in the new millennium, when a plenty publications devoted to such singularities have appeared \cite{sudden2}--\cite{Tomasz}.  
The arising interest to their studies is connected basically with two reasons.
The recent discovery of the cosmic acceleration \cite{cosmic} 
has stimulated the elaboration of dark energy models, responsable for such a phenomenon (see e.g. for review 
\cite{dark}). Remarkably in some of these models the sudden singularities arise quite naturally.
Another source of the interest to sudden singularities is the development of brane models \cite{Shtanov,Shtanov1,sudden9},
where also singularity of this kind arise naturally (sometimes the singularities, arising in the brane models are called ``quiescent'' \cite{Shtanov}). .  

In the investigations devoted to sudden singularities one can distinguish three main topics. First of them deals with the question of  the compatibility of the models, possessing soft singularities with observational data \cite{sudden6,Mar,tach1,Tomasz}.
The second direction is connected with the study of quantum effects \cite{Shtanov1,sudden8,Haro,Kiefer4,quantum,Kiefer2,quantum1,Barbaoza,Calderon}. Here one can 
see two subdirections: the study of quantum corrections to effective Friedmann equation, which can eliminate classical singualrities or, at least, change their form \cite{Shtanov,sudden8,Haro} and the study of solutions of the Wheeler-DeWitt equation for the quantum state of the universe in the presence of sudden singularities \cite{Kiefer4,quantum,Kiefer2,quantum1,Barbaoza}. 
The third direction is connected with the opportunity of the crossing of sudden singularities in classical cosmology 
\cite{Lazkoz,Lazkoz3,Lazkoz1,tach2,quantum1}.

A particular feature of the sudden future singularities is their softness 
\cite{Lazkoz}. As the Christoffel symbols depend only on the first
derivative of the scale factor, they are regular at these singularities.
Hence, the geodesics are well behaved and they can cross the singularity 
\cite{Lazkoz}. One can argue that the particles crossing the singularity
will generate the geometry of the spacetime, providing in such a way a soft rebirth  of the universe after the
singularity crossing \cite{tach2}. Note that the opportunity of crossing of some kind of cosmological 
singularities were noticed already in the early paper by Tipler \cite{Tipler}.
Rather a close idea of integrable singularities in black holes, which can give origin to a cosmogenesis 
was recently put forward in \cite{Lukash,Lukash1}.
Besides, the results of papers \cite{Lazkoz,Lazkoz3} were generalized for the case of 
general (non-Friedmann) universes in papers \cite{Cotsakis1,Cotsakis}. For this purpose was used the formalism of the quasi-isotropic expansion of the solutions of the Einstein equations 
near cosmological singularities, which has been first proposed in \cite{khalat}. (For some further developments of this formalism, see \cite{quasi-isot}).

The peculiarity  of the sudden future singularities makes them to be a good tool for studying some general features of the 
general relativity, in particular the relations between classical and quantum gravity and cosmology. These relations is the main topic 
of the present review. We shall also dwell on  another aspect of general relativity, which from our point of view is a little bit underestimated. It is the fact the that requirement of the self-consistency of the system of laws of general relativity and particle 
physics, or, in other words, of the system of Einstein equations and of the equations describing the state or the motion of non-gravitational 
matter can induce some interesting transformations in the state of matter. Such transformations sometimes occur when the universe passes 
through the soft singularities, though there are some examples of such transformations which can be observed in the absence of singularities too. We shall consider some of such examples. 

Generally, this review is devoted to three interrelated topics, which are connected in some way with the soft future singularities in 
cosmology - these are the problem of crossing of such singularities in classical cosmology, the relations between classical and quantum 
treatments of cosmological singularities and the changes of state of matter, induced by cosmological singularities or other geometrical 
irregularities in the framework of general relativity. The structure of the paper is the following. In the second section we shall give a brief and convenient classification of the future singularities, following the paper \cite{Odin}.  
In the section 3 we present the classification of the types of singularities from the point of view of the finite objects, which approach these singularities. In the section 4 we introduce the toy tachyon model \cite{tach0} and shall discuss its basic properties.   
The section 5 is devoted to the cosmological model based on the mixture of the anti-Chaplygin gas and to the paradox of soft singularity crossing \cite{Paradox}. In sixth section we consider 
again the paradox of the soft singularity crossing in the presence of dust and shall discuss 
its possible resolution by introducing some transformation of matter \cite{Paradox1}. 
In seventh section we shall give another example of the transformation of the Lagrangian of 
a scalar field due to its interaction with geometry, while its potential is not smooth \cite{cusped}.  
The eighth section is devoted to the study of classical dynamics of the cosmological model with a scalar field whose potential is inversely proportional to the field, while in the ninth we study its quantum dynamics. 
The section 10 is devoted to attempts to apply the formalsm of the Wheeler-DeWitt equation to the study of tachyon and pseudotachyon cosmological models.  
In eleventh section we study Friedmann equations modified by quantum corrections and possible influence of these corrections on soft cosmological singularities
\cite{sudden8},\cite{Haro}.
The section 12 is devoted to developing of such notions as density matrix of the universe, quantum consistency and interplay between geometry and matter in quantum cosmology. 
In the section 13 we consider singularities arising in some braneworld models, while the section 14 contains some concluding remarks. 

\section{Classification of future cosmological singularities}
In this section we  shall present rather a conveninent classification of the future cosmological singularities, following the paper \cite{Odin}.
We shall consider a flat Friedmann universe with the metric 
\begin{equation}
ds^2 = dt^2 - a^2(t)dl^2,
\label{Fried}
\end{equation}
where $a(t)$ is the cosmological radius (scale factor) and $dl^2$ is the spatial interval.
We shall choose such a normalization of the gravitational constant which provides the follwoing form of the first Friedmann equation
is
\begin{equation}
H^2 = \rho,
\label{Fried1}
\end{equation}
where 
\begin{equation}
H \equiv \frac{\dot{a}}{a} 
\label{Hubble}
\end{equation}
is the Hubble parameter and $\rho$ is the energy density of the universe. 
The second Friedmann or Raychaudhuri equation is 
\begin{equation}
\frac{\ddot{a}}{a} = -\frac12(\rho+3p),
\label{Fried2}
\end{equation}
where $p$ is the pressure.
The energy conservation equation looks as 
\begin{equation}
\dot{\rho} + 3H(\rho+p) = 0.
\label{cons}
\end{equation}

We shall write down also the expressions for the non-vanishing components of the Riemann-Christoffel curvature tensor,
defined as \cite{Land}
\begin{equation}
R^{i}_{\ klm} = \frac{\partial \Gamma^{i}_{km}}{\partial x^{i}} - \frac{\partial \Gamma^{i}_{kl}}{\partial x^m}+\Gamma^{i}_{nl}\Gamma^{n}_{km}-\Gamma^i_{nm}\Gamma^n_{kl}.
\label{Riem0}
\end{equation}
These nonvanishing components are 
\begin{equation}
R^{\alpha}_{\ t\beta t} = -\frac{\ddot{a}}{a}\delta^{\alpha}_{\beta} = (-\dot{H}+H^2)\delta^{\alpha}_{\beta},  
\label{Riem1}
\end{equation}
where $\alpha,\beta$ are spatial indices;
\begin{equation}
R^{1}_{\ 212} = R^{1}_{\ 313} = R^{2}_{\ 323} = \dot{a}^2 
\label{Riem2}
\end{equation}
and the corresponding components arising from symmetry.

The singularities of the type I are the so called Big Rip singularities \cite{Star-Rip,Rip}.
A type I singularity arises at some finite moment of the cosmic time $t \rightarrow t_{BR}$, when $a \rightarrow \infty, \dot{a} \rightarrow \infty, H \rightarrow \infty, \rho \rightarrow \infty, |p| \rightarrow \infty$. These singularities are present in the model, where the cosmological 
evolution is driven by the so called phantom matter \cite{phantom}, when $p < 0, |p| > \rho$ or, in other words the equation of state parameter $w \equiv \frac{p}{\rho} < -1$.\\
The conditions of arising and avoiding of such singularities were studied in detail in papers 
\cite{Gonzalez-Diaz,Mariam3,Mariam}.

The singularities of the type II are characterized by the following behaviour of the cosmological parameters: at finite 
interval of time $t = t_{II}$ a universe arrives with the finite values of the cosmological radius of the time derivative of the cosmological radius, of the Hubble parameter and of the energy density $t \rightarrow t_{II}, a \rightarrow a_{II}, \dot{a} \rightarrow \dot{a}_{II},
H \rightarrow H_{II}, \rho \rightarrow \rho_{II}$ while the acceleration of the universe and the first time derivative of the Hubble parameter
tends to minus infinity $\ddot{a} \rightarrow -\infty, \dot{H} \rightarrow -\infty$ and the pressure tends to plus infinity $p \rightarrow \infty$. A particular case of the type II singularity is the Big Brake singularity, first found in paper \cite{tach0}.
At this singularity, the time derivative of the cosmological radius, the Hubble variable and the energy density are equal exactly to zero.
\\

The type III singularities are the singularities occuring when the cosmological radius is finite, while its time derivative, the Hubble 
variable, the energy density and the pressure are divergent. The examples of such singularities were considered, for example in \cite{sudden2,Cannata}.\\

The more soft singularities are the singularities of the type IV, at finite value of the cosmological factor both the energy density and the pressure tend to zero and only the higher derivatives of the Hubble parameter $H$ diverge. These singularities sometimes are  called Big Separation singularities. 

In paper \cite{w-sing} the type V singularities were added to the scheme proposed in \cite{Odin}. These are the singularities which like 
the singularities of the type IV have the pressure and energy density tending  to zero, but the higher time derivatives of the Hubble parameter 
are regular and only the barotropic index (equation of state parameter) $w$ is singular.\\

Sometimes the traditional Big Bang and Big Crunch singularities are called type 0 singularities, (see \cite{Mar-Kon}).

In this review we shall mainly speak about type II singularities and their comparison with type 0 singularities.

\section{The type of the singularity from the point of view of finite size objects, which approach these singularities}
In this section we shall present the classification of singularities,
based on the point of view of finite size objects, which approach these 
singularities. In principle, finite size objects could be destroyed while
passing through the singularity due to the occurring infinite tidal forces.
A strong curvature singularity is defined by the requirement that an
extended finite object is crushed to zero volume by tidal forces. We give
below Tipler's \cite{Tipler} and Kr\'{o}lak's \cite{Krolak} definitions of
strong curvature singularities together with the relative necessary and
sufficient conditions.

First of all we shall write down the geodesics deviation equation. 
If $u^i$ are four-velocities of test particles and $\eta^i$ is a four-vector separating two 
spatially close geodesics, then the dynamics of this vector is given by the equation \cite{Land}
\begin{equation}
\frac{D^2\eta^i}{ds^2} = R^{i}_{\ klm}u^ku^l\eta^m,
\label{deviation}
\end{equation}
where $D$ is the covariant derivative along a geodesics.
In the case of a flat Friedmann universe (\ref{Fried}) for the geodesics of particles, having 
zero spatial velocities (i.e $u^{\alpha} = 0, u^t = 1$) Eq. (\ref{deviation}) acquires, taking into account 
Eq. (\ref{Riem1}), a simple form
\begin{equation}
\ddot{\eta}^{\alpha} = R^{\alpha}_{\ tt\beta}\eta^{\beta} = \frac{\ddot{a}}{a}\eta^{\alpha}.  
\label{deviation1}
\end{equation}
Looking at the above equation one can see that approaching a singularity, chracterized by an infinite value 
of the deceleration, we experience an infinite force, stopping the farther increase of the separation of geodesics, while 
geodesics themselves can be quite regular if the the velocity of expansion $\dot{a}$ is regular.     

According to Tipler's definition if every volume element, defined by three
linearly independent, vorticity-free, geodesic deviation vectors along every
causal geodesic through a point $P$, vanishes, a strong curvature singularity
is encountered at the respective point $P$ \cite{Tipler}, \cite{Lazkoz}. The
necessary and sufficient condition for a causal geodesic to run into a
strong singularity at $\lambda _{s}$ ($\lambda $ is affine parameter of the
curve) \cite{ClarkeKrolak} is that the double integral%
\begin{equation}
\int_{0}^{\lambda }d\lambda ^{\prime }\int_{0}^{\lambda ^{\prime }}d\lambda
^{\prime \prime }\left\vert R_{\,\,ajb}^{i}u^{a}u^{b}\right\vert
\label{cond1}
\end{equation}%
diverges as $\lambda \rightarrow \lambda _{s}$. A similar condition is valid
for lightlike geodesics, with $R_{\,\,ajb}^{i}u^{a}u^{b}$ replacing $%
R_{\,ab}u^{a}u^{b}$ in the double integral.

Kr\'{o}lak's definition is less restrictive. A future-endless,
future-incomplete null (timelike) geodesic $\gamma $ is said to terminate in
the future at a strong curvature singularity if, for each point $P\in \gamma 
$, the expansion of every future-directed congruence of null (timelike)
geodesics emanating from $P$ and containing $\gamma $ becomes negative
somewhere on $\gamma $ \cite{Krolak}, \cite{GenStrong}. The necessary and
sufficient condition for a causal geodesic to run into a strong singularity
at $\lambda _{s}$ \cite{ClarkeKrolak} is that the integral%
\begin{equation}
\int_{0}^{\lambda }d\lambda ^{\prime }\left\vert
R_{\,\,ajb}^{i}u^{a}u^{b}\right\vert  \label{cond2}
\end{equation}%
diverges as $\lambda \rightarrow \lambda _{s}$. Again, a similar condition
is valid for lightlike geodesics, with $R_{\,\,ajb}^{i}u^{a}u^{b}$ replacing 
$R_{\,ab}u^{a}u^{b}$ in the integral.

We conclude this section by mentioning that the singularities of the types 0 and I are strong 
and the singularities of the types II, IV and V are 
week
according
to both the definitions (Tipler's and Kr\'{o}lak's ones), while the type III singularities are strong with respect 
to Kr\'{o}lak's definition and week with respect to Tipler's definition \cite{Mar-Kon}.  

The weekness of the type II singularities, which we shall study in some details in the next sections of the present review, according to both the definitions, means that although the tidal forces become infinite, the finite objects are not necessarily crushed when reaching the singularity.

\section{The tachyon cosmological model with the trigonometric potential} 
The tachyon field, born in the context of the string theory \cite{Sen},
provides an example of matter having a large enough negative pressure to produce an acceleration of the expansion rate of the universe. Such a field is today considered as one of the possible candidates for the role of dark energy and, also for this reason, in the recent years it has been intensively studied. The tachyon models represent a subclass of the models with non-standard kinetic terms \cite{k-ess}, which descend from the Born-Infeld model, invented already in thirties \cite{Born}.
Before considering the model with the trigonometric potential \cite{tach0}, possessing the Big Brake singularity, we write down the general formulae of the tachyon cosmology.

The Lagrangian of the tachyon field $T$ is 
\begin{equation}
L = -V(T)\sqrt{1-g^{\mu\nu}T_{,\mu}T_{,\nu}}
\label{tach}
\end{equation}
or, for the spatially homogeneous tachyon field, 
\begin{equation}
L =-V(T)\sqrt{1-\dot{T}^2}.
\label{tach1}
\end{equation}
The energy density and the pressure of this field are respectively 
\begin{equation}
\rho = \frac{V(T)}{\sqrt{1-\dot{T}^2}}
\label{tach2}
\end{equation}   
and
\begin{equation}
p = -V(T)\sqrt{1-\dot{T}^2},
\label{tach3}
\end{equation} 
while the field equation is 
\begin{equation}
\frac{\ddot{T}}{1-\dot{T}^2} + 3H\dot{T} +\frac{V_{,T}}{V(T)} = 0.
\label{KGT}
\end{equation}

We shall introduce also the pseudo-tachyon field with the Lagrangian \cite{tach0} 
\begin{equation}
L = W(T)\sqrt{\dot{T}^2-1}
\label{pseud}
\end{equation}
and with the energy density 
\begin{equation}
\rho = \frac{W(T)}{\sqrt{\dot{T}^2-1}}
\label{pseud1} 
\end{equation}
and the pressure
\begin{equation}
p = W(T)\sqrt{\dot{T}^2-1}.
\label{pseud2}
\end{equation}
The Klein-Gordon equation for the pseudo-tachyon field is 
\begin{equation}
\frac{\ddot{T}}{1-\dot{T}^2} + 3H\dot{T} +\frac{W_{,T}}{W(T)} = 0.
\label{KGTp}
\end{equation}

We shall also write down the equations for the time derivative of the Hubble parameter in the tachyon and pseudo-tachyons models:
\begin{equation}
\dot{H} = -\frac32\frac{V(T)\dot{T}^2}{\sqrt{1-\dot{T}^2}},
\label{Hdott}  
\end{equation}
\begin{equation}
\dot{H} = -\frac32\frac{W(T)\dot{T}^2}{\sqrt{\dot{T}^2-1}}.
\label{Hdotpt}  
\end{equation}
We see that the Hubble parameter in both these models is decreasing. 

Note that for the case when the potential of the tachyon field $V(T)$ is a constant, 
the cosmological model with this tachyon coincides with the cosmological model with the 
Chaplygin gas \cite{FKS}. The Chaplygin gas is the perfect fluid, satisfying the equation of state 
\begin{equation}
p = -\frac{A}{\rho}, \ A > 0.
\label{Chap}
\end{equation}
The cosmological model based on the Chaplygin gas was introduced in \cite{we-Chap} and has acquired some popularity
as a unified model of dark matter and dark energy \cite{unified}.
Analogously, the pseudo-tachyon model with the constant potential 
coincides with the model with  a perfect fluid, whose equation of state is .
\begin{equation}
p = +\frac{A}{\rho}, \ A > 0.
\label{anti-Chap}
\end{equation}
This fluid can be called ``anti-Chaplygin gas''. The corresponding model was introduced in \cite{tach0} and we shall 
come back to it later. Curiosly, similar equation of motion arises in the theory of wiggly strings \cite{wiggly}.

Now we shall study  a very particular tachyon potential depending on the trigonometrical functions which was 
suggested in the paper \cite{tach0}. Its form is 
\begin{eqnarray}
&&V(T) = \frac{\Lambda}{\sin^2\frac32\sqrt{\Lambda(1+k)}T}\nonumber \\
&&\times \sqrt{1-(1+k)\cos^2\frac32\sqrt{\Lambda(1+k)}T},
\label{poten}
\end{eqnarray} 
where $\Lambda$ is a positive constant and $k$ is a parameter, which is chosen in the interval
$-1 < k < 1$. The case of the positive values of the parameter $k$ is  especially interesting.
The set of possible cosmological evolutions, is graphically presented in Figure 1, which is the phase portrait 
of our dynamical system, where the ordinate $s$ is the time derivative of the tachyon field $T$: $s \equiv \dot{T}$. 
\begin{figure}[t]
\includegraphics{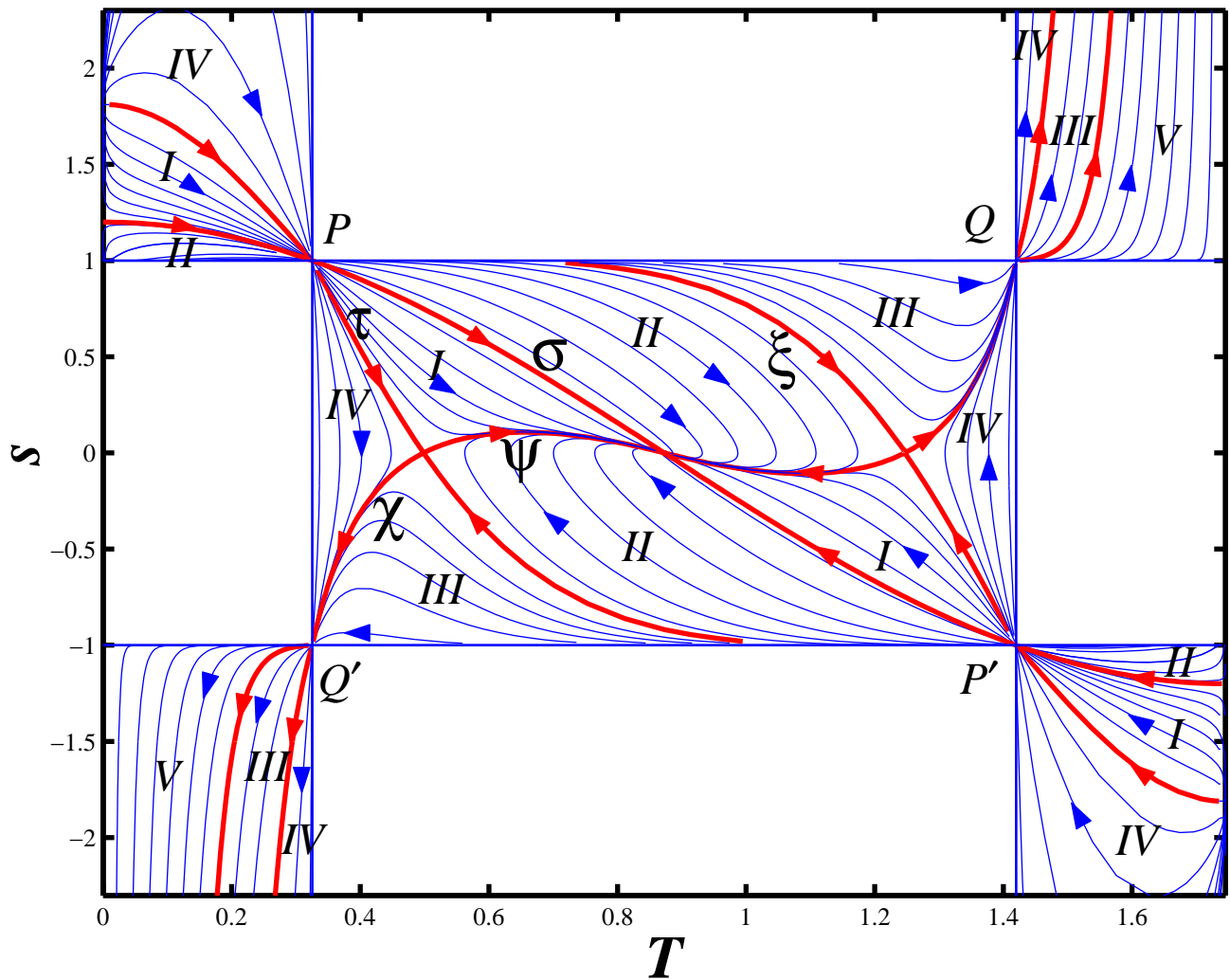} 
\caption{(Color online) Phase portrait evolution for $k > 0$ ($k = 0.44$)}
\label{Fig1}
\end{figure}

The origin of the potential (\ref{poten}) is the following one: let us consider a flat Friedmann universe filled with two 
fluids, one of which is a cosmological constant with the equation of state $p = -\rho = -\Lambda$ and the second one is
a barotropic fluid with the equation of state $p = k\rho$. The Friedmann equation for such a model is exactly solvable and gives 
\begin{equation}
H(t)=\sqrt{\Lambda}\coth\frac{3\sqrt{\Lambda}(k+1)t}{2}.
\label{simple-evol}
\end{equation}
Then using the standard technique of the reconstruction of potentials, which was mainly used for the minimally coupled scalar field \cite{reconstruct}, but was easily generalized for the 
cases of non-minimally coupled fields \cite{nonmin,reconstruct-nonmin}  and for tachyons \cite{Feinstein,Padman,tach0,serena} we 
obtain the expression (\ref{poten}). It is necessary to emphasize that the dynamics of the tachyon model with the potential (\ref{poten}) is much richer than the dynamics of the two fluid model with the unique cosmological evolution given by the expression (\ref{simple-evol}).   
In paper \cite{tach0} both the cases $k \leq 0$ and 
$k > 0$ were considered. The case $k > 0$ is of a particular interest, because it reveals two unusual phenomena:
a self-transformation of the tachyon into a pseudotachyon field and the appearance of the 
Big Brake cosmological singularity.  

Let us discuss briefly the classical dynamics of the model with the trigonometric potential 
for the case $k > 0$. It is easy to see that the potential (\ref{poten}) is well defined at 
$T_3 \leq T \leq T_4$, where
\begin{equation}
T_3 = \frac{2}{3\sqrt{(1+k)\Lambda}} {\rm arccos} \frac{1}{\sqrt{1+k}},
\label{T3}
\end{equation} 
\begin{equation}
T_4 = \frac{2}{3\sqrt{(1+k)\Lambda}} \left(\pi - {\rm arccos} \frac{1}{\sqrt{1+k}}\right).
\label{T4}
\end{equation} 
In turn, the kinetic term $\sqrt{1-\dot{T}^2}$ is well defined at $-1 \leq s \leq 1$.
In other words, the Lagrangian (\ref{tach1}) with the potential (\ref{poten}) is well defined 
inside the rectangle (see Fig. 1). The analysis of the dynamics of the equation of motion 
of the tachyon (\ref{KGT}) and of the Friedmann equations shows that a part of the trajectories 
end their evolution in the attractive node with the coordinates $T_0 = \frac{\pi}{3\sqrt{(1+k)\Lambda}}, s_0 = 0$, which describes an infinite de Sitter expansion. The upper and lower 
borders of the rectangle $s =1, s = -1$, excluding the corner points, are the standard Big Bang cosmological singularities, while left and right borders $T = T_3$ and $T = T_4$ repel 
the trajectories. However, another part of the trajectories goes towards the corner points 
$(T=T_3, s = -1)$ and $(T=T_4, s = 1)$. These points are regular points from the point of view of the equations of motion of the corresponding dynamical system and besides, 
the direct calculation shows that there are no cosmological singularities there.
Thus, there is no reason which prevents further evolution of the universe through these points.
Indeed, one can see also that the equations of motion and their solutions can be continued 
into the vertical stripes (see Fig. 1). However, to reproduce these equations of motion in the stripes as Euler-Lagrange equations, we should substitute the tachyon Lagrangian 
(\ref{tach1})  by the preudotachyon Lagrangian (\ref{pseud1}) with the potential 
\begin{eqnarray}
&&W(T) = \frac{\Lambda}{\sin^2\frac32\sqrt{\Lambda(1+k)}T}\nonumber \\
&&\times \sqrt{(1+k)\cos^2\frac32\sqrt{\Lambda(1+k)}T-1}\   . 
\label{pseud-poten}
\end{eqnarray} 

Thus, we have seen already the first unusual phenomenon - the self-transformation of the 
tachyon into the pseudotachyon field. Now, the question arises: what happens with the universe after the ``crossing the corner'' and the transformation of the tachyon into the pseudotachyon ?     The analysis of equations of motion carried out in paper \cite{tach0}
shows that the universe in a finite moment of time $t = t_{BB}$ encounter the singularity,
which is characterized by the following values of  cosmological parameters:
\begin{eqnarray}
&&a(t_{BB}) = a_{BB} < \infty,\nonumber \\
&&\dot{a}(t_{BB}) = 0, \nonumber \\
&&\ddot{a}(t)  \rightarrow -\infty,\  {\rm at}\ t \rightarrow t_{BB},\nonumber \\
&&T(t_{BB}) = T_{BB} > 0 \ {\rm (in\  lower\  left\  strip)},\nonumber \\
&&s(t)  \rightarrow -\infty,\  {\rm at}\ t \rightarrow t_{BB},\ {\rm (in\  lower\  left\  strip)}\ \nonumber \\
&&\rho(t_{BB}) = 0, \nonumber \\
&&p(t) \rightarrow +\infty,\ {\rm at}\ t \rightarrow t_{BB}.
\label{BigBrake}
\end{eqnarray}
This singularity was called Big Brake singularity \cite{tach0}. Obviously, it enters into 
the class II of singularities, according to the classification suggested in paper \cite{Odin}
and recapitulated in Sec. 2 of the present review. 
  
Now, it is interesting to confront the prediction of this, a little bit artificial, but rather rich model 
with the observational data coming from the luminosity-redshift relation from Supernovae of type Ia. Such an attempt was undertaken in paper \cite{tach1}, where the set of supernovae
studied in \cite{Wood-Vasey} was used. The strategy was the following : there were scanned the pairs of present values of the tachyon field and of its time derivative (points in phase space) and then they were propagated backwards in time, comparing corresponding luminosity distance - redshift curves with the observational data from SNIa. Then, those pairs of values which appeared to be compatible with the data were chosen as initial conditions 
for the future cosmological evolution. Though the constraints imposed by the data were 
rather severe, both evolutions took place: one very similar to $\Lambda$CDM and ending 
in an exponential (de Sitter) expansion; another with  the transformation of the tachyon into the pseudotachyon and the successive running towards the Big Brake singularity. It was found that a larger value of the model parameter $k$ enhances the 
probability to evolve into a Big Brake. The time intervals until the future encounter with the Big Brake
were calculated and were found to be compatible with the present age of the universe \cite{tach1}. 

The next question, which arises, is the fate of the universe after the encounter with the Big Brake singularity. As was already told above, this singularity is very soft and the geodesics can 
be continued across it. Then the matter, passing through the Big Brake singularity reconstructs 
the spacetime. This process was studied in some detail in paper \cite{tach2}.
The analysis of the equation of motion for the universe approaching the Big Brake singularity gives the following expressions for the basic quantities:
\begin{equation}
T = T_{BB} + \left(\frac{4}{3W(T_{BB})}\right)^{1/3}(t_{BB}-t)^{1/3},
\label{T-BB}
\end{equation}
\begin{equation}
s = - \left(\frac{4}{81W(T_{BB})}\right)^{1/3}(t_{BB}-t)^{-2/3},
\label{s-BB}
\end{equation}
\begin{equation}
a=a_{BB} -\frac34 a_{BB}\left(\frac{9W^2(T_{BB})}{2}\right)^{1/3}(t_{BB}-t)^{4/3},
\label{a-BB}
\end{equation}
\begin{equation}
\dot{a}=a_{BB}\left(\frac{9W^2(T_{BB})}{2}\right)^{1/3}(t_{BB}-t)^{1/3},
\label{adot-BB}
\end{equation}
\begin{equation}
H=\left(\frac{9W^2(T_{BB})}{2}\right)^{1/3}(t_{BB}-t)^{1/3}.
\label{H-BB}
\end{equation}

The expressions (\ref{T-BB})--(\ref{H-BB}) can be continued into the region where 
$t > t_{BB}$, which amounts to crossing the Big Bang singularity. Only the expression 
for $s$ is singular at $t=t_{BB}$, but this singularity is integrable and not dangerous.

Upon reaching the Big Brake, it is impossible  for the system to stop there because the 
infinite deceleration leads to the decrease of the scale factor. This is because after the 
Big Brake crossing the time derivative of the cosmological radius (\ref{adot-BB}) and of the Hubble variable (\ref{H-BB}) change their signs. The expansion is then followed by a contraction. Corresponding to given initial conditions, the values of $T_{BB}, t_{BB}$ and 
$a_{BB}$ were found numerically. Then the numerical integration of the equations of motion 
describes the contraction of the universe, culminating in the encounter with the Big Crunch singularity. Curiously, the time intervals between the Big Brake and Big Crunch singularities 
practically do not depend on the initial conditions and are equal approximately to $0.3\times
10^{9}$ yrs \cite{tach2}. 

 Now, the next question arises: what happens if we consider a little bit more complicated 
 model, adding to the tachyon matter some quantity of  dust-like matter ? Obviously, in this case instead of the Big Brake singularity the universe will encounter a soft type II singularity 
 of a more general kind. Namely, due to the presence of dust, the energy density of the expanding universe cannot vanish and, hence, at the moment when the universe experiences 
 an infinite deceleration its expansion should continue. This implies the appearance of some 
 kind of contradictions, which can be resolved by transformation of the pseudotachyon field 
 into another kind of Born-Infeld like field. The corresponding problem was considered in detail in papers \cite{Paradox,Paradox1}. The first of these papers was devoted to a more simple model, based on mixture of the anti-Chaplygin gas with dust. The next section will be devoted to this model.   
  
\section{The cosmological model based on the mixture of the anti-Chaplygin gas and the paradox of soft singularity crossing}

The anti-Chaplygin gas with the equation of state (\ref{anti-Chap}) is one of the simplest 
cosmological models revealing the Big Brake singularity \cite{tach0}. Indeed, combining 
the equation of state (\ref{anti-Chap}) with the energy conservation equation (\ref{cons}),
one obtains immediately
\begin{equation}
\rho = \sqrt{\frac{B}{a^6}-A},
\label{anti-Chap1}
\end{equation}
where $B$ is a positive constant, characterizing the initial condition. Then, when in the process of the cosmological expansion the cosmological radius $a$ arrives to the critical 
value 
\begin{equation}
a_S = \left(\frac{B}{A}\right)^{1/6} 
\label{anti-Chap2}
\end{equation}
the energy density of the universe vanishes while the pressure tends to infinity.
Thus, the universe encounters the Big Brake singularity. Then, it begins contraction
culminating in the encounter with the Big Crunch singularity. 
 
Now, let us see what happens if we add some amount of dust with the energy density
\begin{equation}
\rho_m = \frac{\rho_{0}}{a^3},
\label{dust}
\end{equation}
where $\rho_{0}$ is a positive constant. 
 In this case the traversability of the
singularity seems to be obstructed. The main reason for this is that while
the energy density of the anti-Chaplygin gas
vanishes at the singularity, the energy density of the matter component does
not, leaving the Hubble parameter at the singularity with a finite value.
Then some kind of the paradox arises: if the universe continues its
expansion, and if the equation of state of the component of matter,
responsible for the appearance of the soft singularity (in the simplest
case, the anti-Chaplygin gas) is unchanged, then the expression for the
energy density of this component becomes imaginary, which is unacceptable.
The situation looks rather strange: indeed, the model, including dust should 
be in some sense more regular, that that, containing only such an exotic fluid as 
the anti-Chaplygin gas. Thus, if the model, based on the pure anti-Chaplygin gas has a 
traversable Big Brake singularity, than the more general singularity arising in the model,
based on the mixture of the anti-Chaplygin gas and dust should also be transversable.
 
A possible way of resolution of this paradox, based on use of the distributional cosmological quantities was suggested in paper \cite{Paradox}. Let us suppose that at the moment of the 
crossing of the soft cosmological singularity the expansion of the universe with the Hubble parameter $H$ is abruptly substituted by the cosmological contraction with the Hubble parameter $-H$. In this case, the value of the cosmological radius $a$ begins decreasing 
and the expression (\ref{anti-Chap1}) for the energy density just like the corresponding 
expression for the pressure remain well defined. The first Friedmann equation    
(\ref{Fried1}) and the energy conservation equation (\ref{cons}) remain also intact. 
A problem, however, arises with the second Friedmann equation (\ref{Fried2}).
Let us rewrite this equation in the form 
\begin{equation}
\dot{H} = -\frac32 (\rho + p).
\label{Fried3}
\end{equation}
If the Hubble parameter abruptly changes sign at the moment $t = t_S$  that means that it contains the term 
\begin{equation}
H(t) = H_S(\theta(t_S-t) - \theta(t-t_S),
\label{theta}
\end{equation}
where $\theta(x)$ is the Heaviside theta function. The derivative of the theta function is equal in the distributional sense to the Dirac delta function (see e.g. \cite{Gelfand}). Hence, the left-hand side of Eq. (\ref{Fried3}) contains the Dirac delta function. 
Now, let us discuss in more detail the expressions for the Hubble parameter and its time
derivative in the vicinity of the singularity. The leading terms of the expression for $H(t)$ are
\begin{eqnarray}
H(t) &=&H_{S}sgn(t_{S}-t)  \nonumber \\
&&+\sqrt{\frac{3A}{2H_{S}a_{S}^{4}}}sgn(t_{S}-t)\sqrt{|t_{S}-t|}~,
\label{Hubble-sing2}
\end{eqnarray}%
where $sgn(x) \equiv \theta(x) - \theta(-x)$.
Then
\begin{equation}
\dot{H}=-2H_{S}\delta (t_{S}-t)-\sqrt{\frac{3A}{8H_{S}a_{S}^{4}}}\frac{%
sgn(t_{S}-t)}{\sqrt{|t_{S}-t|}}~.  \label{Hubble-der-sing}
\end{equation}%
Naturally, the $\delta $-term in $\dot{H}$ arises because of the jump in $H$%
, as the expansion of the universe is followed by a contraction. To restore
the validity of the second Friedmann equation (\ref{Fried3})   we shall add a singular $\delta $
-term to the pressure of the anti-Chaplygin gas, which will acquire the form 
\begin{equation}
p_{ACh}=\sqrt{\frac{A}{6H_{S}|t_{S}-t|}}+\frac{4}{3}H_{S}\delta (t_{S}-t)~.
\label{pressure-new}
\end{equation}%
The equation of state  of the anti-Chaplygin gas is
preserved, if we also modify the expression for its energy density: 
\begin{equation}
\rho _{ACh}=\frac{A}{\sqrt{\frac{A}{6H_{S}|t_{S}-t|}}+\frac{4}{3}H_{S}\delta
(t_{S}-t)}~.  \label{en-den-new}
\end{equation}
The last expression should be understood in the sense of the composition of 
distributions (see Appendix A of the paper \cite{Paradox} and references therein). 

In order to prove that $p_{ACh}$ and $\rho _{ACh}$ represent a
self-consistent solution of the system of cosmological equations, we shall
use the following distributional identities: 
\begin{eqnarray}
\left[ sgn\left( \tau \right) g\left( \left\vert \tau \right\vert \right) %
\right] \delta \left( \tau \right) &=&0\ ,  \label{proposition1} \\
\left[ f\left( \tau \right) +C\delta \left( \tau \right) \right] ^{-1}
&=&f^{-1}\left( \tau \right) \ ,  \label{proposition2} \\
\frac{d}{d\tau }\left[ f\left( \tau \right) +C\delta \left( \tau \right) %
\right] ^{-1} &=&\frac{d}{d\tau }f^{-1}\left( \tau \right) \ .
\label{proposition3}
\end{eqnarray}%
Here $g\left( \left\vert \tau \right\vert \right) $ is bounded on every
finite interval, $f\left( \tau \right) >0$ and $C>0$ is a constant. These
identities were proven in paper \cite{Paradox}, where was used the approach to the product and the composition of distributions developed in papers \cite{product}.

Due to Eqs. (\ref{proposition2})-(\ref{proposition3}), $\rho _{ACh}$
vanishes at the singularity while still being continuous. The first term in
the expression for the pressure (\ref{pressure-new}) diverges at the
singularity. Therefore the addition of a Dirac delta term, which is not
changing the value of $p_{ACh}$ at any $\tau \neq 0$ (i.e. $t\neq t_{S}$)
does not look too drastic and might be considered as a some kind of
renormalization.

To prove that the first and the second Friedmann equations and the continuity equation are satisfied
we must only investigate those terms, appearing in the field equations,
which contain Dirac $\delta$-functions.  First, we
check the continuity equation for the anti-Chaplygin gas. Due to the
identities (\ref{proposition2})-(\ref{proposition3}), the $\delta \left(
\tau \right) $-terms occurring in $\rho _{ACh}$ and $\dot{\rho}_{ACh}$ could
be dropped. We keep them however in order to have the equation of state
explicitly satisfied. Then the $\delta \left( \tau \right) $-term appearing
in $3Hp_{ACh}$ vanishes, because the Hubble parameter changes sign at the
singularity (see Eq. (\ref{proposition1})).

The $\delta \left( \tau \right) $-term appearing in $\rho _{ACh}$ does not
affect the Friedmann equation due to the identity (\ref{proposition2}).
Finally, the $\delta $-term arising in the time derivative of the Hubble
parameter in the left-hand side of the Raychaudhuri equation is compensated
by the conveniently chosen $\delta $-term in the right-hand side of Eq. (\ref%
{pressure-new}). 

However, the mathematically self-consistent scenario, based on the use of generalized functions and on the abrupt change of the expansion into a contraction, looks rather counter-intuitive from the physical point of view. Such a behaviour can 
be compared with the absolutely elastic bounce of a ball from a rigid wall, as studied in classical mechanics. In the latter case the velocity and the momentum of the ball change their 
direction abruptly. Hence, an infinite force acts from the wall onto the ball during an infinitely small interval of time. 

In reality, the absolutely elastic bounce is an idealization of a process of finite time-span during which inelastic deformations of the ball and of the wall occur. Thus, the continuity of the kinematics of the act of bounce implies a more complex and realistic description of the dynamical process of interaction between the ball and the wall. It is reasonable to think that something similar occurs also in the models, including  dust and an anti-Chaplygin gas or a tachyon. The smoothing of the process of a transition from an expanding to a contracting phase should include some (temporary) geometrically induced change of the equation of state of matter or of the form of the Lagrangian. We know that such changes do exist in cosmology. 
In the tachyon model \cite{tach0}, there was the tachyon-pseudotachyon transformation   driven by the continuity of the cosmological evolution. In a cosmological model with the phantom filed with a cusped potential 
\cite{cusped,cusped1}, the transformations between phantom and standard scalar field were considered. Thus, it is quite natural that the process of crossing of the soft singularity should imply similar transformations.
     
However, now the situation is more complicated. It is not enough to require the continuity of evolution of the cosmological radius and of the Hubble parameter. It is necessary also to accept some hypothesis concerning the fate of the change of the equation of state of matter or of the form of the Lagrangian. This problem will be considered in the next section. 

\section{Paradox of soft singularity crossing and its resolution due to transformations of matter}
The strategy of the analysis of the the problem of soft singularity crossing in this section is the following \cite{Paradox1}.
First, we shall consider the model with the anti-Chaplygin gas and dust. We shall require a minimality of the change of the form of the dependence of the energy density and of the pressure, compatible with the continuation of the expansion while crossing the soft singularity. Such a requirement will bring us to the substitution of the anti-Chaplygin gas with the Chaplygin gas with a negative energy density.
(Note, that in another context the Chaplygin gas with a negative energy density was considered in paper \cite{Khalat}).
Then we shall consider the cosmological model based on the pseudotachyon field with a constant potential and dust. It is known that the energy-momentum tensor for such a pseudotachyon field coincides with that of the anti-Chaplygin gas (this fact relating the Chaplygin gas and the tachyon field with a constant potential was found in paper \cite{FKS}).  
Thus, we would like to derive the form of the transformation of the pseudotachyon Lagrangian using its kinship with the anti-Chaplygin gas. As a result, we shall come to  a new type of the Lagrangian, belonging to the ``Born-Infeld family''. Finally, we shall extend the found form of transformation of the pseudotachyon field for the case of the field with the trigonometric potential.         

As follows from Eqs. (\ref{anti-Chap}) and (\ref{anti-Chap1})  
the pressure of the anti-Chaplygin
gas 
\begin{equation}
p = \frac{A}{\sqrt{\frac{B}{a^6}-A}}
\label{pressure-anti}
\end{equation}
and it tends to $+\infty$ when the universe approaches the soft singularity, when the cosmological radius $a \rightarrow a_S$ (see Eq. (\ref{anti-Chap2})). If we would like to continue the expansion into the region $a > a_S$, while changing minimally the equation of state we can require 
\begin{equation}
p = \frac{A}{\sqrt{|\frac{B}{a^6}-A|}},
\label{pressure-new0}
\end{equation}
or, in other words, 
\begin{equation}
p = \frac{A}{\sqrt{A-\frac{B}{a^6}}},\ {\rm for}\ a > a_S.
\label{pressure-new1}
\end{equation}
We see that in some ``generalized sense'' we conserve the continuity of the pressure crossing the soft singularity. It passes $+\infty$ conserving its sign. Combining the expression (\ref{pressure-new1}) with the energy conservation law (\ref{cons}) we obtain
\begin{equation}
\rho = -\sqrt{A-\frac{B}{a^6}}\ {\rm for}\ a > a_S.
\label{energy-new}
\end{equation}

Thus, the energy density is also continuous passing through its vanishing value and changing its sign. It is easy to see that the energy density (\ref{energy-new}) and the pressure (\ref{pressure-new1}) satisfy the Chaplygin gas equation of state
\begin{equation}
p = -\frac{A}{\rho}.
\label{Chaplygin}
\end{equation}
Thus, we have seen the transformation of the anti-Chaplygin gas into the Chaplygin gas with a negative energy density. The Friedmann equation after the crossing the singularity is     
\begin{equation}
H^{2}=\frac{\rho _{m,0}}{a^{3}}-\sqrt{A}\sqrt{1-\left( \frac{a_{S}}{a}%
\right) ^{6}}.  
\label{Friedmann2}
\end{equation}%
It follows immediately from Eq. (\ref{Friedmann2}) that after achieving the point of maximal expansion $a = a_{\rm max}$, where  
\begin{equation}
a_{\rm max }=\left( \frac{\rho _{m,0}^{2}}{A}+a_{S}^{6}\right) ^{1/6},
\label{astar}
\end{equation}%
The universe begins contracting. When the contracting universe arrives to $a=a_S$ it again stumbles upon a soft singularity and the Chaplygin gas transforms itself into the anti-Chaplygin gas with positive energy density and the contraction continues until hitting 
the Big Crunch singularity. 

Remember that  in the preceding section and in paper \cite{Paradox}, the process was described when the 
universe passed from the expanding to the collapsing phase instantaneously at the
singularity causing a jump in the Hubble parameter. Here we showed that the
continuos transition to the collapsing phase is possible if the equation of
state of anti-Chaplygin gas has a some kind of  a ``phase transition'' at the singularity.

When the potential of the pseudotachyon field is constant, $W(T) =W_0$, then the energy density (\ref{pseud1}) and the pressure (\ref{pseud2}) satisfy the anti-Chaplygin gas equation of 
state (\ref{anti-Chap}) with 
\begin{equation}
A = W_0^2.
\label{WA}
\end{equation}  
Solving the equation of motion for the pseudotachyon field (\ref{KGTp})
with $W(T) = W_0$ one finds
\begin{equation}
\dot{T}^{2}=\frac{1}{1-\left( \frac{a}{a_{S}}\right) ^{6}}  \label{Tdot1}
\end{equation}%
and we see that the soft singularity arises at $a = a_S$, when $\dot{T}^2 \rightarrow +\infty$. 

Now, we would like to change the Lagrangian (\ref{pseud}) in such a way that the new Lagrangian gives us the energy density and the pressure satisfying the Chaplygin gas equation with a negative energy density. It is easy to check that the Lagrangian 
\begin{equation}
L = W_0\sqrt{\dot{T}^2+1}
\label{BI-new}
\end{equation}
giving 
\begin{equation}
p = W_0\sqrt{\dot{T}^2+1}
\label{pressure-BI}
\end{equation}
and 
\begin{equation}
\rho = -\frac{W_0}{\sqrt{\dot{T}^2+1}}
\label{energy-BI}
\end{equation}
is what we are looking for. 

Note, that the energy density and the pressure, passing through the singularity are continuous in the same sense in which they were continuos in the case of the anti-Chaplygin gas. Thus, we have introduced a new type of the Born-Infeld field, which can be called ``anti-tachyon''. Generally, its Lagrangian is 
\begin{equation}
L = W(T) \sqrt{\dot{T}^2+1}
\label{Lagr-new}
\end{equation}
and the equation of motion is 
\begin{equation}
\frac{\ddot{T}}{\dot{T}^2+1}+3H\dot{T} - \frac{W_{,T}}{W} = 0.
\label{KG-BI}
\end{equation}
For the case $W(T) = W_0$, the solution of equation (\ref{KG-BI}) is
\begin{equation}
\dot{T}^{2}=\frac{1}{\left( \frac{a}{a_{S}}\right) ^{6}-1},  \label{Tdot2}
\end{equation}%
and the energy density evolves as%
\begin{equation}
\rho _{T}=-W_0\sqrt{1-\left( \frac{a_{S}}{a}\right) ^{6}}
\label{rho-BI}
\end{equation}
and the evolution of the universe repeats that for the model with the anti-Chaplygin gas and dust.

Let us emphasize once again that to the transformation from the anti-Chaplygin gas 
to the Chaplygin gas corresponds to the transition from the prseudotachyon field with the Lagrangian (\ref{pseud}) to the new type of the Born-Infeld field, which we can call
``quasi-tachyon field'' with the Lagrangian (\ref{Lagr-new}).

Now, we shall consider the case of the toy model with the trigonometric potential in the presence of dust. We have seen that  
the Born-Infeld type pseudotachyon field runs into a soft Big Brake singularity with
the expansion of the universe in this model.  However, what happens in the presence of dust
component? Does the universe still run into soft singularity? 

To answer this question rewrite Eq. (\ref{KGTp}) as 
\begin{equation}
\ddot{T} = (\dot{T}^2-1)\left(3H\dot{T}+\frac{W_{,T}}{W}\right).
\label{KG1}
\end{equation}
It is easy to see that in the left lower and in the right upper stripes (see Fig. 1), where 
the trajectories describe the expansion of the universe after the transformation of the 
tachyon into the pseudotachyon field, the signs of $\ddot{T}$, of $\dot{T}$ and of the 
term $\frac{W_{,T}}{W}$ coincide. The detailed analysis based on this fact was carried out 
in paper \cite{tach0} and led to the conclusion that the universe encounters the singularity
as $T \rightarrow T_S$ ($T_S > 0$ or $T_S > T_{max}$) , $|\dot{T}| \rightarrow \infty$.
The presence of dust cannot alter this effect because it increases the influence of the term 
$3H\dot{T}$, and hence, accelerates the encounter with the singularity. 

However, the presence of dust changes in an essential way the time dependence of the tachyon field close to the singularity. As it was shown in \cite{tach2} (see also the Sec. 4 of the present paper)
\begin{equation}
T = T_{BB} + \left(\frac{4}{3W(T_{BB})}\right)^{1/3}(t_{BB}-t)^{1/3},
\label{old-asymp}
\end{equation}
while in the presence of dust one has 
\begin{equation}
T = T_S + \sqrt{\frac{2}{3H_S}}\sqrt{t_S-t},
\label{new-asymp}
\end{equation}
where $H_S$ is the nonvanishing value of the Hubble parameter given by 
\begin{equation}
H_S = \sqrt{\frac{\rho_{m,0}}{a_S^3}}.
\label{H-S}
\end{equation}
It is easy to see that the smooth continuation of the expression (\ref{new-asymp}) is impossible in contrast to the situation without dust (\ref{old-asymp}) considered in 
\cite{tach2}. 

Thus, the presence of dust is responsible for the appearance of similar paradoxes in both the anti-Chaplygin gas and tachyon models. 

In the vicinity of the soft singularity, it is the ``friction'' term $3H\dot{T}$ in the 
equation of motion (\ref{KGTp}) , which dominates over the potential term $\frac{W_{,T}}{W}$, hence, the dependence of $W(T)$ is not essential and a pseudotachyon field   approaching     
this singularity behaves like one with a constant potential. Thus, it is quite reasonable to suppose that  crossing of the soft singularity the pseudotachyon transforms itself into the 
quasi-tachyon with the Lagrangian (\ref{Lagr-new}).

Now, we can analyze the dynamics 
of the anti-tachyon field, driven by the equation of motion (\ref{KG-BI}) and by the Friedmann 
equation, where the right-hand side includes the dust contribution and the anti-tachyon 
energy density 
\begin{equation}
\rho = -\frac{W(T)}{\sqrt{\dot{T}^2+1}}.
\label{anti-energy}
\end{equation}
It is convenient to consider the processes developing in the left lower strip of the phase diagram of the model to facilitate the comparison with the earlier studies of the dynamics 
of the tachyon model without dust, undertaken in papers \cite{tach0,tach2}.    

One can see that the relative sign of the term with the second derivative $\ddot{T}$ 
with respect to the friction term $3H\dot{T}$ are oppostite for the pseudotachyons and anti-tachyons. That means that after the crossing of the soft singularity the time derivative $\dot{T}$
is growing and its absolute value is diminishing. At the same time the value of the field $T$ 
is diminishing and the value of the potential $W(T)$ is growing. That means that the absolute 
value of the negative contribution to the energy density of the universe coming from the quasi-tachyon is growing while the energy density of the dust is diminishing due to the expansion of the universe. At some moment this process brings us  to the vanishing value of the general energy density and we arrive to the point of  maximal expansion of the universe. After that the expansion is replaced by the contraction and the Hubble variable changes  sign. The change of  sign of the friction term $3H\dot{T}$ implies the diminishing of the value of $\dot{T}$ and 
at some finite moment of time the universe again encounters the soft singularity when 
$\dot{T} \rightarrow -\infty$. Passing this singularity the quasi-tachyon transforms itself back to 
the pseudotachyon and the relative sign of the terms with the second and first time derivatives
in the equation of motion for this field changes once again. After that the time derivative 
of the pseudotachyon field begins growing and the universe continues its contraction until 
it encounters with the Big Crunch singularity.      

It was shown in paper \cite{tach2} that for the case of the purely tachyon model with the trigonometric potential  the encounter of the universe with the Big Crunch singularity occurs at $T = 0$ and $\dot{T} = -\sqrt{\frac{1+k}{k}}$. One can show that the presence of dust does not change these values. Indeed, let us consider the behavior of the  
 pseudotachyon field when $T \rightarrow 0,\ \dot{T} \rightarrow - \sqrt{\frac{1+k}{k}}$.
 It follows from the expressions (\ref{pseud1}) and (\ref{pseud2}) that the ratio between the pressure 
 and the energy density behaves as 
 \begin{equation}
 \frac{p}{\rho} = \dot{T}^2 -1 \rightarrow \frac{1}{k},
 \label{barotropic}
 \end{equation}
 i.e. in the vicinity of the Big Crunch singularity the pseudotachyon field behaves as a barotropic 
 fluid with the the equation of state parameter $\frac{1}{k} > 1$. That means that the energy 
 density of the pseudotachyon field is growing as 
 \begin{equation}
 \rho \sim \frac{1}{a^{3(1+\frac1k)}}
\label{rad-sing}
\end{equation}
as $a \rightarrow 0$, i.e. much more rapidly than the dust energy density. Thus, one can 
neglect the contribution of the dust in this regime of approaching the Big Crunch singularity
and the description of the evolution of the universe to this point coincides with that of the 
pure tachyon model \cite{tach2}.

\section{The transformations of the Lagrangian of a scalar field with a cusped potential}  
It is well known that the cosmological observations gives as a best fit for the equation of state parameter $w = frac{p}{\rho}$ a value which is slightly inferior with respect to $-1$ 
(see, e.g. \cite{phantom-observe}). The corresponding type of dark energy was called 
``phantom'' matter \cite{phantom}. Wanting to realize such a dark matter using a minimally coupled scalar field, one has to introduce for the latter a negative kinetic term. Thus, its Lagrangian has the form
\begin{equation}
L = -\frac{1}{2}g^{\mu\nu}\phi_{,\mu}\phi_{,\nu} - V(\phi).
\label{phantom} 
\end{equation}
Some observations also indicate that the value of the equation of state parameter at some moment in the past has crossed the value $w =-1$, corresponding to the cosmological constant. Such a phenomenon has received the name of  ``phantom divide line crossing''
\cite{divide}. A minimally coupled scalar field, describing non phantom dark energy 
has a kinetic term with the positive sign. So, it looks natural, to use two scalar fields, a phantom field with the negative kinetic term and a standard one to describe the phantom divide line crossing \cite{two-field}. Another posssible way of the phantom divide line crossing,  using s scalar field nonminimally coupled to gravity was considered in papers \cite{nonmin}. 

However, in papers \cite{cusped,cusped1} it was shown that considering potentials with cusps and 
choosing some particular initial conditions, one can describe the phenomenon of the phantom divide line crossing in the model with one minimally coupled scalar field. Curiously, 
a passage through the maximum point of the evolution of the Hubble parameter  implies 
the change of sign of the kinetic term. Though a cosmological singularity is absent in these cases, this phenomenon is a close relative of those, considered in the preceding sections, because here also we stumble upon some transformation of matter properties, induced by a change of geometry. One can add that in this aspect the phenomenon of the phantom divide line crossing is the close analog of the transformation between the tachyon and pseudo-tachyon model, in the tachyon model with the trigonometric potential, described in Sec. 4. 
Here, we shall present a brief sketch of the ideas, described in papers \cite{cusped,cusped1}, emphasizing the analogy and the differences between different geometrically induced matter transformations.   

We begin with a simple mechanical analog: a particle moving in a potential  
with a cusp \cite{cusped}. 
Let us consider a one-dimensional problem of a classical point
particle moving in the potential
\begin{equation}
V(x) = \frac{V_0}{(1+x^{2/3})^2},
\label{classical}
\end{equation}
where $V_0 > 0$.
The equation of motion is 
\begin{equation}
\ddot{x} -\frac{4V_0}{3(1+x^{2/3})^3 x^{1/3}} = 0.
\label{classical1}
\end{equation}
We consider three classes of possible motions characterized 
by the value of the energy
$E$. The first class consists of the motions when $E < V_0$. 
Apparently, the particle 
with $x < 0, \dot{x} > 0$ or with $x > 0, \dot{x} < 0$ cannot 
reach the point $x = 0$ 
and stops at the points $\mp
\left(\sqrt{\frac{V_0}{E}}-1)\right)^{3/2}$ 
respectively.

The second class includes the trajectories when $E > V_0$. 
In this case the particle crosses the point $x = 0$ with nonvanishing velocity. 

If we have a fine tuning such that $E = V_0$, we encounter an exceptional case. 
Now the  trajectory satisfying Eq. (\ref{classical1})
in the vicinity of the point $x = 0$ can behave as 
\begin{equation}
x = C(t_0-t)^{3/2},
\label{classical2}
\end{equation}
where 
\begin{equation}
C = \pm \left(\frac{16V_0}{9}\right)^{3/4}
\label{C-define}
\end{equation}
and $t \leq t_0$. 
It is easy to see that independently of the sign of $C$ in Eq. (\ref{C-define}) 
the signs of the particle coordinate $x$ and of its velocity $\dot{x}$ are opposite and hence, the particle can arrive in finite time to the point of the cusp of the potential 
$x = 0$. 

Another solution reads as 
\begin{equation}
x = C(t-t_0)^{3/2},
\label{classical4}
\end{equation}
where $t \geq t_0$.
This solution describes the particle going away from the point $x = 0$. 
Thus, we can combine the branches of the solutions (\ref{classical2}) and (\ref{classical4}) 
in four different manners and there is no way to choose if the particle arriving to the point 
$x=0$ should go back or should pass the cusp of the potential (\ref{classical}). It can stop at the top as well. Such a ``degenerate'' behaviour of the particle in this third case is connected 
with the fact that  this trajectory is the separatrix between two one-parameter families of trajectories described above. 
At the moment there is not yet any strict analogy between this separatrix and the cosmological evolution describing the phantom divide line. In order to establish a closer analogy  and to understand what is the crucial difference between mechanical consideration and  general relativistic one, we 
can try to introduce a friction term into the Newton equation (\ref{classical1})
\begin{equation}
\ddot{x} + \gamma\dot{x}-\frac{4V_0}{3(1+x^{2/3})^3 x^{1/3}} = 0.
\label{classical5}
\end{equation} 
It is easy to check that, if the friction coefficient $\gamma$ 
is a constant, one does not have 
a qualitative change in respect to the discussion above. 
Let us asuume for $\gamma$ the dependence 
\begin{equation}
\gamma = 3\sqrt{\frac{\dot{x}^2}{2}+V(x)}.
\label{gamma} 
\end{equation}
then
\begin{equation}
\dot{\gamma} = -\frac{3}{2}\dot{x}^2
\label{gammadot}
\end{equation}
and 
\begin{equation}
\ddot{\gamma} = -3\ddot{x}\dot{x}
\label{gammaddot}
\end{equation}
just like in the cosmological case, where the role of the friction coefficient is played by the Hubble parameter. 
The trajectory arriving to the cusp with vanishing velocity is still 
described by 
the solution (\ref{classical2}). Consider the particle coming to the
cusp 
from the left 
($C < 0$. It is easy to see that the value of $\dot{\gamma}$ at the 
moment $t_0$ tends to zero,
while its second derivative $\ddot{\gamma}$ given by Eq. (\ref{gammaddot}) is 
\begin{equation}
\ddot{\gamma}(t_0) = \frac98 C^2 > 0.
\label{gammaddot1}
\end{equation}
Thus, it looks like  the friction coefficient $\gamma$ reaches its
minimum 
value at $t = t_0$. 
Let us suppose now that the particle is coming back to the left from
the 
cusp and its motion is described by Eq. (\ref{classical4}) with
negative $C$. 
A simple check shows that 
in this case 
\begin{equation}
\ddot{\gamma}(t_0) = -\frac98 C^2 < 0.
\label{gammaddot2}
\end{equation}
Thus, from the point of view of the subsequent evolution this point
looks 
as a maximum 
for the function $\gamma(t)$. In fact, it means simply that the second 
derivative of 
the friction coefficient has a jump at the point $t = t_0$.
It is easy to check that if instead of choosing the motion to the
left, 
we shall move forward our particle to the right from the cusp ($C>0$), 
the sign of $\ddot{\gamma}(t_0)$ remains negative as in
Eq. (\ref{gammaddot2}) 
and hence we have the jump of this second derivative again. If one
would 
like to avoid 
this jump, one should try to change the sign in Eq. (\ref{gammaddot}). 
To implement it in a self-consistent way one can substitute 
Eq. (\ref{gamma}) by 
\begin{equation}
\gamma = 3\sqrt{-\frac{\dot{x}^2}{2}+V(x)}
\label{gamma1} 
\end{equation}
and Eq. (\ref{classical5}) by 
\begin{equation}
\ddot{x} + \gamma\dot{x}+\frac{4V_0}{3(1+x^{2/3})^3 x^{1/3}} = 0.
\label{classical51}
\end{equation} 
In fact, it is exactly that what 
happens automatically in cosmology, when we change the sign of the kinetic energy term for the scalar field, 
crossing the phantom divide line. Naturally, in cosmology 
the role of $\gamma$ is played by the Hubble variable $H$.  
The jump of the second derivative of the friction coefficient $\gamma$ corresponds to the divergence of the third time derivative of the Hubble variable, which represents some kind of very soft cosmological singularity. 

Thus, one seems to confront the problem of choosing between two alternatives: 1) to encounter a weak singularity in the spacetime geometry; 2) to change the sign of the kinetic term for matter field.  
We have pursued the second alternative insofar as
we privilege the smoothness of spacetime geometry and consider equations of motion for matter as less 
fundamental than the Einstein equations. 

Now, we would like to  say that the potential, considered in papers \cite{cusped,cusped1}
had the general structure 
\begin{equation}
V(\phi) = \frac{1}{A+B\phi^{2/3}}. 
\label{cusped-pot}
\end{equation}
The origin of this structure is the following: one considers the power law expansion of the universe, it is well-known that such 
an expansion could be provided by an exponential potential \cite{Matarrese}. Then one can represent the Friedmann equation for the evolution of the scale factor of the universe as a second-order linear differential equation, where the potential is reperesented as a function of the time parameter \cite{Yurov}. This equation has two independente solutions: one of them is the power-law expansion and other corresponds to an evolution driven by a phantom matter. The linear combination of these two solutions with both nonvanishing coefficients gives an evolution, where a universe crosses the phantom divide line.  
It is impossible to reconstruct the form of the potential as a function of the scalar field, which provides such an evolution explicitly, however, one can study its form around the point where the phantom divide crossing  occur and this form is exactly that of Eq. (\ref{cusped-pot}) \cite{cusped}.

At the end of this section, we would like to say that in the Newtonian mechanics there is rather a realistic example of motion when, the dependence of the distance of time is given by some fractional power \cite{Newton,Newton1}. Indeed, if one consider the motion of a car with a constant power (which is more realistic than the motion with  a constant force, usually presented in textbooks), when the velocity behaves as $t^{1/2}$ and if the initial value 
of the coordinate and of the velocity are equal to zero, when the acceleration behaves as 
$t^{-1/2}$ and at the moment of start is singular. The motion at constant power is an excellent model of drag-car racing \cite{Newton,Newton1}. Its analogy with the cosmology at the presence of sudden singularities was noticed in paper \cite{Cotsakis}.

\section{Classical dynamics of the cosmological model with a scalar field whose potential is inversely proportional to the field}

We have considered earlier the simplest model, possessing a soft cosmological singularity (Big Brake) - the model based on  
the anti-Chaplygin gas. It was noticd that this model is equivalent to the model with the pseudotachyon field with constant potential. Here we would like to study a model, based on a minimally coupled scalar field, which possesses the same evolution as the model based on the anti-Chaplygin gas. Using the standard technique of the reconstruction of potential, the potential of the corresponding scalar field was found in paper \cite{quantum} and it looks like 
\begin{equation}
V(\varphi) = \pm\frac{\sqrt{A}}{2}\left(\sinh 3\varphi -\frac{1}{\sinh 3\varphi}\right).
\label{poten-scal}
\end{equation}
As a matter of fact we have two possible potentials, which differs by the general sign.
We choose the sign ``plus''. Then, let us remember that the Big Brake occurs when the
energy density is equal to zero (the disappearance of the Hubble parameter) and the pressure is positive and infinite (an infinite deceleration). To achieve this condition, in the scalar field model it is necessary to require that the potential is negative and infinite. It
is easy to see from Eq. (\ref{poten-scal}) that this occurs when $\varphi \rightarrow 0$ being positive. Thus, to have the model with the Big Brake singularity we can consider the scalar field with a potential which is a little bit simpler than that from Eq. (\ref{poten-scal}), but still possesses rather a rich dynamics. Namely we shall study the scalar field with the potential
\begin{equation}
V = -\frac{V_0}{\varphi},
\label{poten1}
\end{equation}
where $V_0$ is a positive constant. 
The Klein-Gordon equation for the scalar field with the potential (\ref{poten1}) is
\begin{equation}
\ddot{\varphi} + 3H\dot{\varphi}+\frac{V_{0}}{\varphi^{2}}=0
\label{KG}
\end{equation}
while the first Friedmann equation is
\begin{equation}
H^{2}=\frac{\dot{\varphi }^{2}}{2}-\frac{V_{0}}{\varphi} \hspace{0.1cm}.
\label{Fried20}
\end{equation}
We shall also need the expression for the time derivative of the Hubble parameter, which
can be easily obtained from Eqs. (\ref{KG}) and (\ref{Fried20}):
\begin{equation}
\dot{H}=-\frac{3}{2}\dot{\varphi }^{2} \hspace{0.1cm}.
\label{Hdot}
\end{equation}

Now we shall construct the complete classification of the cosmological evolutions
(trajectories) of our model, using Eqs. (\ref{KG})-(\ref{Hdot}) \cite{quantum1}.

First of all, let us announce briefly the main results of our analysis.
\begin{enumerate}
\item
The transitions between the positive and negative values of the scalar field are impossible.
\item
All the trajectories (cosmological evolutions) with positive values of the scalar field 
begin in the Big Bang singularity, then achieve a point of maximal expansion, then contract and 
end their evolution in the Big Crunch singularity.
\item   
All the trajectories with positive values of the scalar field pass through the point where the value of the scalar field 
is equal to zero. After that the value of the scalar field begin growing. The point $\varphi = 0$ corresponds to a crossing of the soft singularity.
\item 
If the moment when the universe achieves the point of the maximal expansion coincides with the moment of the  crossing of the soft singularity  then the singularity is the Big Brake. 
\item
The evolutions with the negative values of the scalar field belong to two classes - first, an infinite expansion beginning from 
the Big Bang 
and second, the evolutions obtained by the time reversion of those of the first class, which are contracting and end 
in the Big Crunch singularity. 
\end{enumerate}

To prove these results, we begin with the consideration of the universe in the vicinity of the point $\varphi = 0$.
We shall look for the leading term of the field $\varphi$ approaching this point in the form
\begin{equation}
\varphi(t) = \varphi_1(t_S-t)^{\alpha},
\label{lead}
\end{equation}
where $\varphi_1$ and $\alpha$ are positive constants and $t_S$ is the moment of the soft singularity crossing.
The time derivative of the scalar field is now 
\begin{equation}
\dot{\varphi}(t) = \alpha\varphi_1(t_S-t)^{\alpha-1}.
\label{lead1}
\end{equation}
Because of the negativity of the potential (\ref{poten1}) at positive values of $\varphi$, the kinetic term should be stronger than the potential one to satisfy the Friedmann equation (\ref{Fried2}). That implies that $\alpha 
\leq \frac23$.  However, if $\alpha < \frac23$ we can neglect the potential term and remain with the massless scalar field.
It is easy to show considering the Friedmann (\ref{Fried2}) and Klein-Gordon (\ref{KG}) equations that in this case the scalar 
field behaves like $\varphi \sim \ln (t_S-t)$, which is incompatible with the hypothesis of its smallness (\ref{lead}). Thus, 
one remains with the only choice 
\begin{equation}
\alpha = \frac23.
\end{equation}
Then, if the coefficient at the leading term in the kinetic energy is greater than that in the potential, it follows from the Friedmann
equation (\ref{Fried2})  that the Hubble parameter behaves as $(t_S-t)^{-\frac13}$ which is incompatible with Eq. (\ref{Hdot}). Thus, the leading terms of the potential and kinetic energy should cancel each other: 
\begin{equation}
\frac12\alpha^2\varphi_1^2(t_S-t)^{2\alpha-2} =\frac{V_0}{\varphi_1}(t_S-t)^{-\alpha},
\label{compare}
\end{equation}
that for $\alpha = \frac23$ gives 
\begin{equation}
\varphi_1 =\left(\frac{9V_0}{2}\right)^{\frac13}.
\label{varphi1}
\end{equation}
Hence, the leading term for the scalar field in the presence of the soft singularity is 
\begin{equation}
\varphi(t) = \left(\frac{9V_0}{2}\right)^{\frac13}(t_S-t)^{\frac23}.
\label{brake}
\end{equation}
Now, integrating Eq. (\ref{Hdot}) we obtain
\begin{equation}
H(t) = 2\left(\frac{9V_0}{2}\right)^{\frac23}(t_S-t)^{\frac13} +H_S,
\label{Hbrake}
\end{equation}
where $H_S$ is an integration constant giving the value of the Hubble parameter at the moment of the soft singularity crossing.
If this constant is equal to zero, $H_S = 0$, the moment of the maximal expansion of the universe coincides with 
that of the soft singularity crossing and the universe encounters the Big Brake singularity. If $H_S \neq 0$ we have a more 
general type of the soft cosmological singularity where the energy density of the matter in the universe is different from zero.
The sign of $H_S$ can be both, positive or negative, hence, universe can pass through this singularity in the phase of its 
expansion or of its contraction.

The form of the leading term for the scalar field in the vicinity of the moment when $\varphi = 0$ (\ref{brake}) shows that, after passing the zero value, the scalar field begin growing being positive. Thus, it proves the first result from the list
presented above about impossibility of the change of the sign of the scalar field in our model. 

We have already noted that the time derivative of the scalar field had changed the sign crossing the soft singularity.
It cannot change the sign in a non-singular way because the conditions $\dot{\varphi}(t_0) = 0, \varphi(t_0) \neq 0$ are incompatible with the Friedmann  equation (\ref{Fried2}). It is seen from Eq. (\ref{brake}) that before the crossing of the soft singularity  
the time derivative of the scalar field is negative and after its crossing it is positive. The impossibility of the changing 
the sign of the time derivative of the scalar field without the soft singularity crossing implies the inevitability of the approaching of the universe to this soft singularity. Thus, the third result from the list above is proven. 

It is easy to see from Eq. (\ref{Hdot}) that the value of the Hubble parameter is decreasing during all the evolution. 
At the same time, the absolute value of its time derivative (proportional to the time derivative squared of the scalar field) 
is growing after the soft singularity crossing. That means that at some moment the Hubble parameter should change its sign 
becoming negative. The change of the sign of the Hubble parameter is nothing but the passing through the point of the maximal 
expansion of the universe, after which it begin contraction culminating in the encounter with the Big Crunch singularity.
Thus the second result from the list presented above is proven. 

Summing up, we can say that all the cosmological evolutions where the scalar field has positive values have the following structure: they begin in the Big Bang singularity with an infinite positive value of the scalar field and an infinite negative value of its time derivative, then they pass through the soft singularity where the value of the scalar field is equal to zero and where the derivative of the scalar field changes its sign. All the trajectories also pass through the point of the maximal expansion, and this passage trough the point of the maximal expansion can precede or follow the passage trough the soft singularity:
in the case when these two moments coincide ($H_S = 0$) we have the Big Brake singularity (see the result 4 from the list above). Thus, all the evolutions pass through the soft singularity, but only for one of them this singularity has a character of the Big Brake singularity. The family of the trajectories can be parameterized by the value of the Hubble parameter $H_S$ at the moment of the crossing of the soft singularity. There is also another natural parameterization of this family - we can 
characterize a trajectory by the value of the scalar field $\varphi$ at the moment of the maximal expansion of the universe and by the sign of its time derivative at this moment (if the time derivative of the scalar field is negative that means that the 
passing through the point of maximal expansion precedes the passing through the soft singularity and if the sign of this time derivative
is positive, then passage trough the point of  maximal expansion follows the passage through the soft singularity). If at the moment when the universe achieves the point of maximal expansion the value of the scalar field is equal to zero, then it is the exceptional trajectory crossing the Big Brake singularity.

For completeness, we shall say some words about the result 5, concerning the trajectories with the negative values of the scalar field. Now, both the terms in the right-hand side of the Friedmann equation (\ref{Fried2}), potential and kinetic, are positive 
and, hence, the Hubble parameter cannot disappear or change its sign. It can only tends to zero asymptotically while 
both these terms tend asymptotically to zero. Thus, in this case there are two possible regimes: an infinite expansion which begins with the Big Bang singularity and an infinite contraction which culminates in the 
encounter with the Big Crunch singularity. The second regime can be obtained by the time reversal of the first one and vice versa. Let us consider the expansion regime. 
It is easy to check that the scalar field being negative cannot achieve the zero value, because the suggestion 
$\varphi(t) = -\varphi_1(t_0-t)^{\alpha}$, where $\varphi_1 < 0, \alpha > 0$ is incompatible with the equations 
(\ref{Fried2}) and (\ref{Hdot}). Hence,  the potential term is always non-singular and at the birth of the universe from the 
Big Bang singularity the kinetic term dominates and the dynamics is that of the theory with the massless scalar field.
Namely 
\begin{equation}
\varphi(t) = \varphi_0 +\sqrt{\frac29}\ln t,\ \ H(t) = \frac{1}{3t},
\label{neg0}
\end{equation}
where $\varphi_0$ is a constant. At the end of the evolution the Hubble parameter tends to zero, while the time grows indefinitely. That means that both the kinetic and potential terms in the right-hand side of Eq. (\ref{Fried2}) should 
tend to zero. It is possible if the scalar field tends to infinity while its time derivative tends to zero.
The joint analysis of Eqs. (\ref{Fried2}) and (\ref{Hdot}) gives the following results for the asymptotic behavior 
of the scalar field and the Hubble parameter:
\begin{equation}
\varphi(t) =\tilde{\varphi}_0-\left(\frac56\right)^{\frac25}V_0^{\frac15}t^{\frac25},\ \ H(t) = \left(\frac65\right)^{\frac15}V_0^{\frac25}t^{-\frac15},
\label{neg1}
\end{equation}
where $\tilde{\varphi}_0$ is a constant.

\section{The quantum dynamics of the cosmological model with a scalar field whose potential is inversely proportional to the field}

The introduction of the notion of the quantum state of the universe, satisfying the Wheeler-DeWitt equation \cite{DeWitt} has stimulated the diffusion of the hypothesis that in the framework
of quantum cosmology the singularities can disappear in some sense. Namely, the probability of finding of the universe
with the parameters, which correspond to a classical cosmological singularity can be equal to zero (for a recent treatments see
\cite{Kiefer,Kiefer3,Kiefer1}).

In this section we shall study the quantum dynamics of the model, whose classical dynamics was described in the preceding section.
Our presentation follows that of papers \cite{quantum,quantum1}.

As usual, we shall use the canonical formalism and the Wheeler-DeWitt equation \cite{DeWitt}. For this purpose, instead of the Friedmann metric (\ref{Fried}), we shall consider a more general metric, 
\begin{equation}
ds^2 = N^2(t)dt^2 -a^2(t)dl^2,
\label{lapse}
\end{equation}
where $N$ is the so-called lapse function. The action of the Friedmann flat model with the minimally coupled scalar field looks now as 
\begin{equation}
S = \int dt \left(\frac{a^3\dot{\varphi}^2}{2N}-a^3V(\varphi) -\frac{a\dot{a}^2}{N}\right).
\label{action}
\end{equation}
Variating the action (\ref{action}) with respect to $N$ and putting then $N=1$ we come to the standard Friedmann equation. 
Now, introducing the canonical formalism, we define the canonically conjugated momenta as 
\begin{equation}
p_{\varphi} = \frac{a^3\dot{\varphi}}{N}
\label{mom-phi}
\end{equation}
and 
\begin{equation}
p_{a} = -\frac{a\dot{a}}{N}.
\label{mom-a}
\end{equation}
The Hamiltonian is
\begin{equation}
{\cal H} = N\left(-\frac{p_a^2}{4a}+\frac{p_{\varphi}^2}{2a^3}+Va^3\right)
\label{Hamilton}
\end{equation}
and is proportional to the lapse function. The variation of the action with respect to $N$ gives the constraint  
\begin{equation}
-\frac{p_a^2}{4a}+\frac{p_{\varphi}^2}{2a^3}+Va^3 = 0,
\label{constraint}
\end{equation}
and the implementation of the Dirac quantization procedure, i.e. requirement the that constraint eliminates the quantum state 
\cite{Dirac}, 
gives the Wheeler-DeWitt equation
\begin{equation}
\left(-\frac{\hat{p}_a^2}{4a}+\frac{\hat{p}_{\varphi}^2}{2a^3}+Va^3\right)\psi(a,\varphi) = 0.
\label{WDW}
\end{equation}
Here $\psi(a,\phi)$ is the wave function of the universe and the hats over the momenta mean that the functions are substituted by the operators. Introducing the differential operators representing the momenta as
\begin{equation}
\hat p_a \equiv \frac{\partial}{i \partial a},\ \ \hat p_{\varphi} \equiv \frac{\partial}{i\partial \varphi}  
\label{dif}
\end{equation}
and multiplying Eq. (\ref{WDW}) by $a^3$ we obtain the following partial differential equation:
\begin{equation}
\left(\frac{a^2}{4}\frac{\partial^2}{\partial a^2} - \frac12\frac{\partial^2}{\partial \varphi^2} +a^6V\right)\psi(a,\varphi) = 0.
\label{WDW1}
\end{equation}
Finally, for our potential inversely proportional to the scalar field we have
 \begin{equation}
\left(\frac{a^2}{4}\frac{\partial^2}{\partial a^2} - \frac12\frac{\partial^2}{\partial \varphi^2} -\frac{a^6V_0}{\varphi}\right)\psi(a,\varphi) = 0.
\label{WDW2}
\end{equation}
Note that in the equation (\ref{WDW}) and in the subsequent equations we have ignored rather a complicated problem of the choice of the ordering of  noncommuting operators, because the specification of such a choice is not essential for our analysis. 
Moreover, the interpretation of the wave function of the universe is rather an involved question \cite{Venturi,barv,Bar-Kam}. The point is that to choose the measure in the space of the corresponding Hilbert space we should fix a particular gauge condition, eliminating in such a 
way the redundant gauge degrees of freedom and introducing  a temporal dynamics into the model \cite{barv}. We shall not dwell here  on this procedure, assuming generally that 
the cosmological radius $a$ is in some way connected with the chosen time parameter and that the unique physical variable is 
 the scalar field $\varphi$. Then, it is convenient to represent the solution of Eq. (\ref{WDW2}) in the form 
\begin{equation}
\psi(a,\varphi) = \sum_{n=0}^{\infty}C_n(a)\chi_n(a,\varphi),
\label{wavefunction}
\end{equation}
where the functions $\chi_n$ satisfy the equation
\begin{equation}
\left(- \frac12\frac{\partial^2}{\partial \varphi^2} -\frac{a^6V_0}{\varphi}\right)\chi(a,\varphi) = -E_n(a)\chi_n(a,\varphi),
\label{chi}
\end{equation}
while the functions $C_n(a)$ satisfy the equation 
\begin{equation}
\frac{a^2}{4}\frac{\partial^2 C_n(a)}{\partial a^2} = E_n(a)C_n(a),
\label{C}
\end{equation}
where $n=0,1,\ldots$.
Requiring the normalizability of the functions $\chi_n$ on the interval $0 \leq \varphi < \infty$, which, in turn, implies   their non-singular behavior at $\varphi =0$ and $\varphi \rightarrow \infty$, and using the considerations similar to those used in the analysis of the Schr\"odinger equation for the hydrogen-like atoms, one can show that the acceptable values of the functions $E_n$ are 
\begin{equation}
E_n = \frac{V_0a^{12}}{2(n+1)^2},
\label{E}
\end{equation}
while the corresponding eigenfunctions are 
\begin{equation}
\chi_n(a,\varphi) = \varphi \exp\left(-\frac{V_0a^{6}\varphi}{n+1}\right)L_n^1\left(\frac{2V_0a^6\varphi}{n+1}\right),
\label{Laguerre}
\end{equation}
where $L_n^1$ are the associated Laguerre polynomials. 

Rather often the fact that the wave function of the universe disappears at the values of the cosmological parameters
corresponding to some classical singularity is interpreted as an avoidance of such singularity. 
However, in the case of the soft singularity considered in the model at hand, such an interpretation 
does not look too convincing. 
Indeed, one can have a temptation to think that the probability of finding of the universe in the soft singularity state
characterized by the vanishing value of the scalar field is vanishing because the expression for functions  (\ref{Laguerre}) entering into the 
expression for the wave function of the universe (\ref{wavefunction}) is proportional to $\varphi$. However, the wave function 
(\ref{wavefunction}) can hardly have a direct probabilistic interpretation. Instead, one should choose some reasonable time-dependent gauge, identifying some combination of  variables with an effective time parameter, and interpreting other 
variables as physical degrees of freedom \cite{barv}. The definition of the wave function of the universe in terms of these 
physical degrees of freedom is rather an involved question; however, we are in a position to make some semi-qualitative 
considerations. The  reduction of the initial set of variables to the smaller set of physical degrees of freedom implies the appearance of the Faddeev-Popov determinant which as usual is equal to the Poisson bracket of the gauge-fixing condition 
and the constraint \cite{barv}. Let us, for example, choose as a gauge-fixing condition the identification of the new ``physical'' time 
parameter with the Hubble parameter $H$ taken with the negative sign. Such an identification is reasonable, because as it follows from Eq. (\ref{Hdot}) 
the variable $H(t)$ is monotonously decreasing. The volume $a^3$ is the variable canonically conjugated to the Hubble variable. Thus, the Poisson bracket between the gauge-fixing condition $\chi = H - T_{\rm phys}$ and the constraint (\ref{constraint}) includes 
the term proportional to the potential of the scalar field, which is inversely proportional to this field itself. Thus, the singularity in $\varphi$ arising in the Faddeev-Popov determinant can cancel zero, arising in ({\ref{Laguerre}).

 Let us confront this situation with that of the Big Bang and Big Crunch singularities. As it was seen in Sec. III 
such singularities classically arise at infinite values of the scalar field. To provide the normalizability of the wave function  one should have the integral on the values of the scalar field $\varphi$ convergent, when $|\varphi| \rightarrow \infty$. 
That means that, independently of details connected with the gauge choice, not only the wave function of the universe but also  the probability density of scalar field values  should decrease 
rather rapidly when the absolute value of the scalar field is increasing. Thus, in this case, the effect of the quantum avoidance of the classical singularity is present.

\section{The quantum cosmology of the tachyon and the pseudo-tachyon field}
In this section we would like to construct the Hamiltonian formalism for the tachyon and pseudo-tachyon fields. Using the metric (\ref{lapse}), one can see that the contribution of the tachyon field into the action is 
\begin{equation}
S=-\int dt Na^3V(T)\sqrt{1-\frac{\dot{T}^2}{N^2}}.
\label{ac-Ham}
\end{equation} 
The conjugate momentum for $T$ is 
\begin{equation}
p_T = \frac{a^3V\dot{T}}{N\sqrt{1-\frac{\dot{T}^2}{N^2}}}.
\label{pT}
\end{equation}
and so the velocity can be expressed as 
\begin{equation}
\dot{T} = \frac{Np_T}{\sqrt{p_T^2+a^6V^2}}.
\label{vel-mom}
\end{equation}
The Hamiltonian of the tachyon field is now 
\begin{equation}
{\cal H} = N\sqrt{p_T^2+a^6V^2}.
\label{ham-tach}
\end{equation}

Analogously, for the pseudo-tachyon field, we have 
\begin{equation}
p_T = \frac{a^3W\dot{T}}{N\sqrt{\frac{\dot{T}^2}{N^2}-1}},
\label{ps}
\end{equation}
\begin{equation}
\dot{T} = \frac{Np_T}{\sqrt{p_T^2-a^6W^2}}
\label{ps1}
\end{equation}
and
\begin{equation}
{\cal H} = N\sqrt{p_T^2-a^6W^2}.
\label{ham-tach1}
\end{equation}
In what follows it will be convenient  for us to fix the lapse function as $N=1$. 

Now, adding the gravitational part of the Hamiltonians and quantizing the corresponding observables, we obtain the following Wheeler-DeWitt equations for the tachyons 
\begin{equation}
\left(\sqrt{\hat{p}_T^2+a^6V^2} - \frac{a^2\hat{p}_{a}^2}{4}\right)\psi(a,T) = 0
\label{WDWT}
\end{equation}
and for the pseudo-tachyons  
\begin{equation}
\left(\sqrt{\hat{p}_T^2-a^6W^2} - \frac{a^2\hat{p}_{a}^2}{4}\right)\psi(a,T) = 0.
\label{WDWTp}
\end{equation}

The study of the Wheeler-DeWitt equation for the universe filled with a tachyon or a pseudo-tachyon field is rather a difficult task because the Hamiltonian depends non-polynomially on the conjugate momentum of such fields. However, one can come to interesting  conclusions, considering some particular models. 

First of all, let us consider a model with the pseudo-tachyon field having a constant potential. 
In this case the Hamiltonian in Eq. (\ref{WDWTp}) does not depend on the field $T$. Thus, it is more convenient to use the representation of  the quantum state of the universe 
where it depends on the coordinate $a$ and the momentum $p_T$. Then the Wheeler-DeWitt equation will have the following form:
\begin{equation}
\left(\sqrt{p_T^2-a^6W^2} + \frac{a^2}{4}\frac{\partial^2}{\partial a^2}\right)\psi(a,p_T) = 0.
\label{WDWTp1}
\end{equation}
It becomes algebraic in the variable $p_T$. Now, we see that the Hamiltonian is well defined at $p_T^2 \geq a^6W^2$. Looking at the limiting value $p_T^2 = a^6W^2$ and comparing it with the relation (\ref{ps1}) we see that it corresponds to $\dot{T}^2 \rightarrow \infty$, which,
in turn, corresponds to the encounter with the Big Brake singularity as was explained in the section V. The only way to ``neutralize'' the  values of $p_T$, which imply the negativity 
of the expression under the square root in the left-hand side of Eq. (\ref{WDWTp1}), is to require that  the wave function of the universe is such that 
\begin{equation}
 \psi(a,p_T) = 0\ \ {\rm at} \ p_T^2 \leq a^6W^2.
 \label{brake-cond}
 \end{equation}
 The last condition could be considered as a hint on 
 the quantum avoidance of the Big Brake singularity. However, as it was explained in Sec. IV on the example of the scalar field 
model, to speak about the probabilities in the neighborhood of the point where the wave function of the universe vanishes,
it is necessary to realize the procedure of the reduction of the set of variables to a smaller set of physical degrees of freedom. Now, let us suppose that the gauge-fixing condition is chosen in such a way that the role of time is played by a Hubble parameter. In this case the Faddeev-Popov determinant, equal to the Poisson bracket between the gauge-fixing condition 
and the constraint, will be inversely proportional to the expression $\sqrt{\hat{p}_T^2-a^6W^2}$ (see Eq. (\ref{ham-tach1})), which tends to zero at the 
moment of the encounter with the Big Brake singularity. Thus, in the case of a pseudo-tachyon model, just like in the case 
of the cosmological model based on the scalar field, the Faddeev-Popov determinant introduces the singular factor, which 
compensates the vanishing of the wave function of the universe.

 What can we say about the Big Bang and the Big Crunch singularities in this model? 
  It was noticed in the preceding section that at these singularities $\dot{T}^2 = 1$. 
  From the relation (\ref{ps1}) it follows that such values of $\dot{T}$ correspond to 
  $|p_T| \rightarrow \infty$. A general requirement of the normalizability of the wave function of the universe implies the vanishing of $\psi(a,p_T)$ at $p_T \rightarrow \pm \infty$ which 
  signifies the quantum avoidance of the Big Bang and the Big Crunch singularities. 
It is quite natural, because these singularity are not traversable in classical cosmology. 

Now we consider the tachyon cosmological model with the trigonometric potential, whose 
classical dynamics was briefly sketched in the  sections 4 and 6. In this case the Hamiltonian depends on both the tachyon field $T$ and its momentum $p_T$. The dependence of the expression under the square root on  $T$ is more complicated than that on $p_T$. Hence,  it does not make sense to use the representation $\psi(a,p_T)$ instead of $\psi(a,T)$. 
Now, we have under the square root  the second order differential operator $-\frac{\partial^2}{\partial T^2}$, which is positively defined, and the function  $-a^6W^2(T)$, which is negatively defined. The complete expression should not be negative, but what does it mean in our case? It means that we should choose such wave functions for which the quantum average of the 
operator $\hat{p}_T^2-a^6W^2(T)$ is non-negative: 
\begin{eqnarray}
&&\langle \psi |\hat{p}_T^2-a^6W^2(T)|\psi \rangle \nonumber \\ 
&&= \int {\cal D}T \psi^*(a,T) \left(-\frac{\partial^2}{\partial T^2} -a^6W(T)^2\right)\psi(a,T) \geq 0.
\label{non-neg}
\end{eqnarray}
Here the symbol ${\cal D}T$ signifies the integration on the tachyon field $T$ with some measure. 
It is easy to guess that the requirement (\ref{non-neg}) does not imply the disappearance of the wave function $\psi(a,T)$ 
at some range or at some particular values of the tachyon field, and one can always construct a wave 
function which is different from zero everywhere and thus does not show the phenomenon of the quantum avoidance of singularity. However, the forms of the potential $V(T)$ given by Eq. (\ref{poten}) and of the corresponding potential $W(T)$ for the 
pseudo-tachyon field arising in the same model \cite{tach0} are too cumbersome to construct such functions explicitly. Thus,
to illustrate our statement, we shall consider a more simple toy model. 

Let us consider the Hamiltonian 
\begin{equation}
{\cal H} = \sqrt{\hat{p}^2 - V_0x^2},
 \label{toy}
\end{equation}
where $\hat{p}$ is the conjugate momentum of the coordinate $x$ and $V_0$ is some positive constant. Let us choose as a wave function a Gaussian function 
\begin{equation}
\psi(x) = \exp(-\alpha x^2),
\label{Gauss}
\end{equation}
where $\alpha$ is a positive number and we have omitted the normalization factor, which is not essential in the present context.
Then the condition (\ref{non-neg}) will look like 
\begin{eqnarray}
&&\int dx \exp(-\alpha x^2) \left(-\frac{d^2}{dx^2}-V_0x^2\right)\exp(-\alpha x^2) \nonumber \\
&&= \sqrt{\frac{\pi}{2}}\left(\frac34\sqrt{\alpha}-
\frac{V_0}{2\alpha^{\frac32}}\right) \geq 0,
\label{non-neg1}
\end{eqnarray}
which can be easily satisfied if 
\begin{equation}
\alpha \geq \sqrt{\frac23 V_0}.
\label{non-neg2}
\end{equation}

Thus, we have seen that for this very simple model one can always  choose such a quantum state, which does not disappear at any value of the coordinate $x$ and which guarantees the positivity of the quantum average of the operator, which is not generally 
positively defined. Coming back to  our cosmological model we can say that the requirement of the well-definiteness of the 
pseudo-tachyon part of the Hamiltonian operator in the Wheeler-DeWitt equation does not imply the disappearance of the 
wave function of the universe at some values of the variables and thus, does not reveal the effect of the quantum avoidance of the cosmological singularity.   

At the end of this section we would like also to analyze the Big Bang and Big Crunch singularities in the tachyon model with the
trigonometrical potential. As was shown in paper \cite{tach0} the Big Bang singularity can occur in two occasions (the same is true also for the Big Crunch singularity \cite{tach2}) - either $W(T) \rightarrow \infty$ (for example for $T \rightarrow 0$) 
or at $\dot{T}^2 =1, W(T) \neq 0$. One can see from Eqs. (\ref{tach2}) and (\ref{ps}) that when the universe approaches these singularities  the momentum $p_T$ tends to infinity. As was explained before, the wave function of the universe in the momentum representation should vanish at $|p_T| \rightarrow \infty$ and hence,  we have the effect of the quantum avoidance.
  
Finally, summing up the content of the last three sections, devoted to the comparative study of the classical and quantum dynamics in some models with scalar fields and tachyons, revealing soft future singularities, we can make the following remarks.

It was shown that in the tachyon model with the trigonometrical potential \cite{tach0} the wave function of the universe is not obliged to vanish in the range of the variables corresponding to the appearance of the classical Big Brake singularity. 
In a more simple pseudo-tachyon cosmological model the wave function, satisfying the Wheeler-DeWitt equation
and depending on the cosmological radius and the pseudo-tachyon  field, disappears at the Big Brake singularity. 
However, the transition to the wave function depending only on the reduced set of physical degrees of freedom implies 
the appearance of the Faddeev-Popov factor, which is singular and which singularity compensates the terms, responsible 
for the vanishing of the wave function of the universe.
Thus, in both these cases, the effect of the quantum avoidance of the Big Brake singularity is absent. 

In the case of the scalar field model with the potential inversely proportional to this field, all the classical trajectories
pass through a soft singularity (which for one particular trajectory is exactly the Big Brake). The wave function of the universe 
disappears at the vanishing value of the scalar field which classically corresponds to the soft singularity. However, also 
in this case the Faddeev-Popov factor arising at the reduction to the physical degrees of freedom provides nonzero value 
of the probability of finding of the universe at the soft singularity. 

In spite of the fact that we have considered 
some particular scalar field and tachyon-pseudo-tachyon models, our main conclusions were based on rather general properties of 
these models. Indeed, in the case of the scalar field we have used the fact that its potential at the soft singularity should be 
negative and divergent, to provide an infinite positive value of the pressure. In the case of the pseudo-tachyon field both the possible vanishing of the wave function of the universe and its ``re-emergence'' in the process of reduction were connected 
with the general structure of the contribution of such a field into the super-Hamiltonian constraint (\ref{ham-tach1}).   
Note that in the case of the tachyon model with the trigonometric potential, the wave function does not disappear at all. 

On the other hand we have seen that for the Big Bang and Big Crunch singularities not only the  wave functions of the universe but also the corresponding probabilities disappear when the universe is approaching to  the corresponding values of the fields under consideration,
and this fact is also connected with rather general properties of the structure of the Lagrangians of the theories.
Thus, in these cases the effect of quantum avoidance of singularities takes place. 

One can say that there is some kind of a classical - quantum 
correspondence here. The soft singularities are traversable at the classical level (at least for simple homogeneous and isotropic Friedmann models) and the effect of quantum avoidance of singularities is absent. The strong Big Bang and Big Crunch 
singularities cannot be passed by the universe at the classical level, and the study of the Wheeler-DeWitt equation 
indicates the presence of the quantum singularity avoidance effect.  

It would be interesting also to find  examples of the absence of the effect of the quantum avoidance of singularities, 
for the singularities of the Big Bang--Big Crunch type. Note that the interest to the study of the possibility 
of crossing of such singularities is growing and some models treating this phenomenon have been elaborated during last few years \cite{Bars}. 

\section{Friedmann equations modified by quantum corrections and soft cosmological singularities}

As we have already mentioned in the Introduction there are two main directions in the study 
of quantum cosmology of soft future singularities. One is connected with the analysis of the structure of the Wheeler-DeWitt equation and another concentrates of the study of quantum corrections to the Friedmann equations. While in two preceding sections we were studying the Wheeler-DeWitt equation, here we shall dwell on the quantum corrections to the Friedmann equations and on the possible influence of these corrections on the structure of soft singularities. Our presentation will be mainly based on papers \cite{sudden8},\cite{Haro}. 

In paper \cite{sudden8} was considered a cosmological evolution described by 
\begin{equation}
a(t) = \left(\frac{t}{t_s}\right)^{1/2}(a_s-1)+1-\left(1-\frac{t}{t_s}\right)^n,
\label{Barrow10}
\end{equation}
where $t_s$ is the time, where the sudden singularity occurs, $a_s$ is the value 
of the scale factor in this moment and $1<n<2$. The matter responsible for this evolution was not specified. It is easy to see that at the beginning of the evolution (\ref{Barrow10}) the universe passes through the radiation-dominated phase of the expansion, while when $t \rightarrow t_s$ it enters into the singular regime. Then it was supposed that a massive scalar field conformally coupled to gravity is present. 
The general solutions describing behaviour of this scalar field in these two regimes were written down and the requirement of the matching of these conditions at the $a = a_s$ was imposed.      
Then, the solution in the first regime is chosen as 
\begin{equation}
\phi_k(\eta) = \frac{e^{ik\eta}}{\sqrt{2k}},
\label{wave}
\end{equation}
where $\eta$ is the conformal time parameter. The solution in the regime of approaching the soft singularity will be 
\begin{equation} 
\phi_k(\eta) = \xi_{01}e^{i\tilde{\omega}\eta} +\xi_{02}e^{-i\tilde{\omega}\eta},
\label{wave1}
\end{equation}
where 
\begin{equation}
\tilde{\omega} = \sqrt{k^2+m^2H_s^2a_s^4}
\label{tilde-omega}
\end{equation}
and the constants $\xi_{01}$ and $\xi_{02}$ are connected with 
the Bogoliubov coefficients :
\begin{equation} 
\alpha = \sqrt{2\tilde{\omega}}\xi_{01},\ \ \beta=\sqrt{2\tilde{\omega}}\xi_{02}.
\label{Bogol}
\end{equation}
The matching conditions permit to find the Bogoliubov coefficients and the number of created particles for each mode
\begin{equation}
N_k =\beta_k\beta_k^* = \frac14\left(1-\frac{k}{\tilde{\omega}}\right)^2.
\label{part-number}
\end{equation}
The total energy of the created particles 
\begin{equation}
\rho = \int \rho_k d^3k =\pi \int k^2\tilde{\omega}
\left(1-\frac{k}{\tilde{\omega}}\right)^2 dk
\label{tot-en}
\end{equation}
is divergent in the ultraviolet limit. The authors of \cite{sudden8} renormalize the expression 
(\ref{tot-en}) using $n$-wave method \cite{n-wave} and show that the renormalized energy is equal to zero. Thus, they conclude that the quantum phenomena associated with the cosmological dynamics do not change the character of the sudden singularity or prevent its occurrence. Some arguments in favour of the hypothesis that birth of particles of a field which is not conformally invariant cannot  change the Friedmann equation are also developed in 
\cite{sudden8}. 

More detailed analysis of the quantum contributions into energy-momentum tensor and, hence, into the Friedmann equations, was undertaken in paper \cite{Haro}. Here it was noticed that the analysis, presented in paper \cite{sudden8}, is applicable only to situations when the frequency of the field under consideration is varying smoothly. Obviously, it is not case here, because two different phases of evolution are considered and a naive matching
of the value of the field and of its time derivative at the moment of arrival to the singularity, is required. Moreover, the effect of polarization of the vacuum was not taken into account. 
Instead, the authors of the paper \cite{Haro}, use the known expressions for the renormalized 
energy-density and pressure for a massless conformally coupled scalar field \cite{Davies,Haro1}:
\begin{equation}
\rho^{\rm ren} = \frac{1}{480\pi^2}\left(3H^2\dot{H}+H\ddot{H}-\frac12\dot{H}^2\right)+\frac{1}{960\pi^2}H^4,
\label{ren-en} 
\end{equation}
\begin{equation}
p^{\rm ren} = -\frac{1}{1440\pi^2}\left(\ddot{H}+11H^2\dot{H}+6H\ddot{H}+\frac92\dot{H}^2\right)-\frac{1}{960\pi^2}H^4.
\label{ren-pres}
\end{equation}
Then proceeding as in paper \cite{Nojiri} the authors of \cite{Haro} consider the Friedmann 
semiclassical equation 
\begin{equation}
H^2 = \rho +\rho^{\rm ren},
\label{Fried-sem}
\end{equation}
looking for its solution with the form 
\begin{equation}
H(t) = H_s -C\left(1-\frac{t}{t_s}\right)^{n'},
\label{Fried-sem1}
\end{equation}
where $H_s, C$ and $n'$ are unknown parameters. They find, in particular, that
\begin{equation}
n' = n+1.
\label{Fried-sem2}
\end{equation}
Then, since $3 < n' < 4$, it turns out that $\dot{H}$ and $\ddot{H}$ do not diverge at $t = t_s$, which means that, for these kinds of singular solutions, the singularity becomes much milder due to the quantum corrections. In fact, in the absence of the quantum corrections, one can see from Eq. (\ref{Barrow10}) that $\dot{H}$ diverges. 

 \section{Density matrix of the universe, quantum consistency and interplay between geometry and matter in quantum cosmology}

In this section we shall speak about the quantum density matrix of the universe \cite{Bar-Kam-dens,Bar-Kam-dens1,Bar-dens,Bar-Kam-Def,Bar-Kam-Def1} - an approach to quantum cosmology, which permits consideration of mixed quantum states of the universe instead of 
pure ones. Such and approach is based on rather a delicate interplay between geometry and matter and implies existence of essential restrictions on the basic parameters of the theory.
In the framework of this approach as a byproduct arise also some new kinds of soft sudden quantum singularities \cite{Bar-Kam-Def}. 

As is well known, quantum cosmology predicts the initial conditions for the cosmological evolution of the universe, defining its quantum state - the wave function of the universe. The connection between the Euclidean quantum theory and the quantum tunneling is used in both the main approaches to the construction of such a function - the no-boundary prescription 
\cite{HH} and the tunneling one\cite{Vil,other-tun}.
In papers \cite{Bar-Kam-dens,Bar-Kam-dens1} this traditional scheme of quantum cosmology was generalized for the case of fundamental mixed initial quantum states of the universe, in other words instead of wave function of the universe one can consider the density matrix of the universe, possessing some thermodynamical characteristics. Such a mixed state of the universe arises naturally if an instanton with two turning points (surfaces of vanishing external curvature) does exist. (The idea that instead of pure quantum state of the universe one can consider a density matrix of the universe, was suggested already in paper \cite{Page}).    

In turn, an instanton with two turning points arises naturally, if we consider a closed Friedmann universe where two essential ingredients are present: an effective cosmological constant and radiation, which corresponds to the presence of the conformally invariant fields.
The Euclidean Friedmann equation in this case is written as 
\begin{equation}
\frac{\dot{a}^2}{a^2} =
    \frac{1}{a^2} - H^2 -\frac{C}{a^4},
    \label{dens-Fried}
\end{equation}    
where $H^2$ is an effective cosmological constant and the constant $C$ characterizes 
the quantity of the radiation in the universe. The turning points are  
\begin{equation}
a_\pm= \frac{1}{\sqrt{2}H}
    \sqrt{1\pm(1-4CH^2)^{1/2}},\ \ 4CH^2 \leq 1.
    \label{dens-turn}
    \end{equation}
(The same instanton was considered also in paper \cite{Gott}, where the conception of the universe, which gave birth to itself was suggested). 
Fig. 2 gives the picture of the instanton representing the density matrix of the universe 
\begin{figure}[t]
\includegraphics[height=.1\textheight]{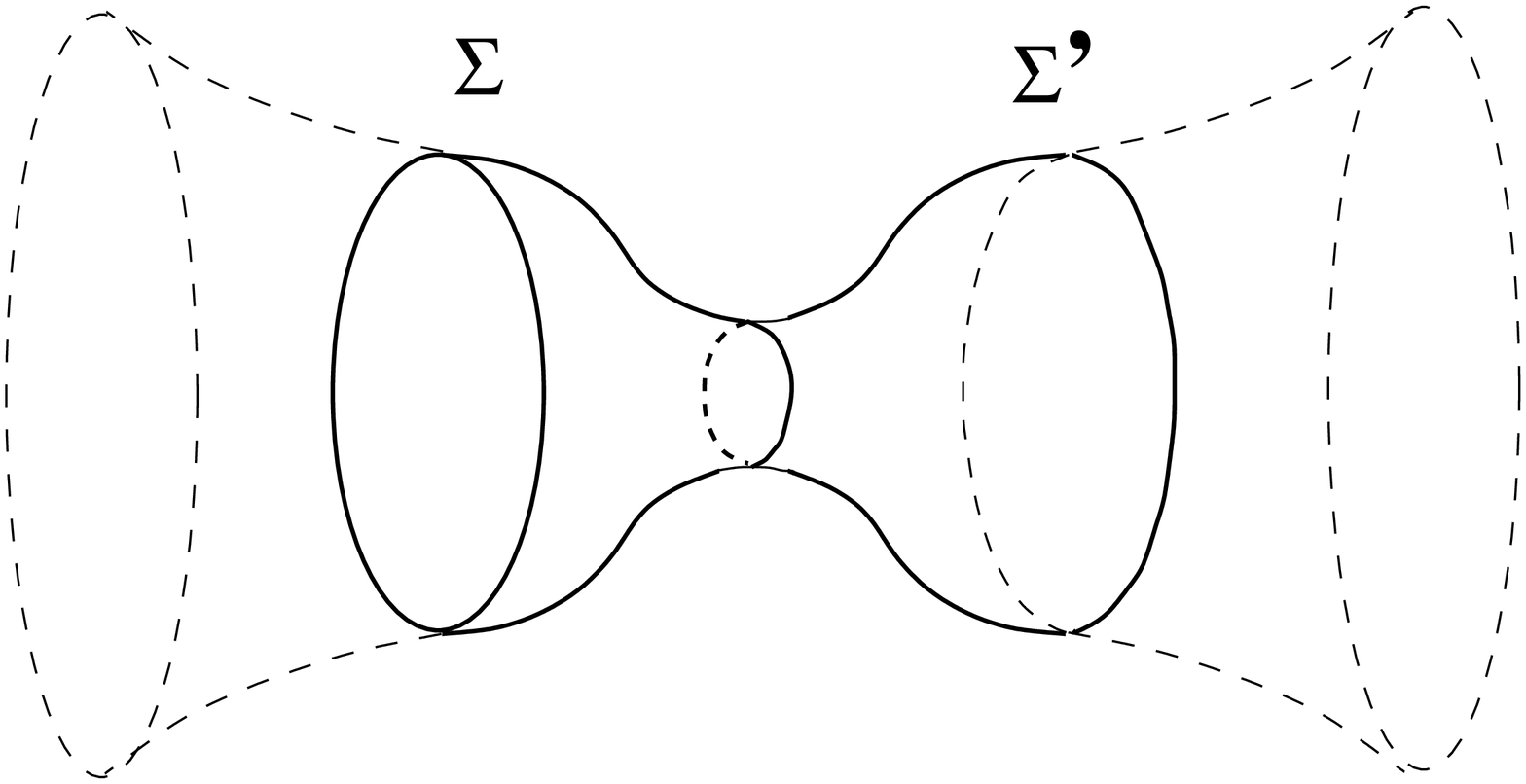} 
\caption{}Picture of instanton representing the density matrix. Dashed lines
depict the Lorentzian Universe nucleating from the instanton at the
minimal surfaces $\Sigma$ and $\Sigma'$.
\label{Fig2}
\end{figure}
For the pure quantum state \cite{HH}  the instanton bridge between
$\Sigma$ and $\Sigma'$ breaks down (see Fig.3). However, the
radiation stress tensor prevents these half instantons from closure.
\begin{figure}
\includegraphics[height=.1\textheight]{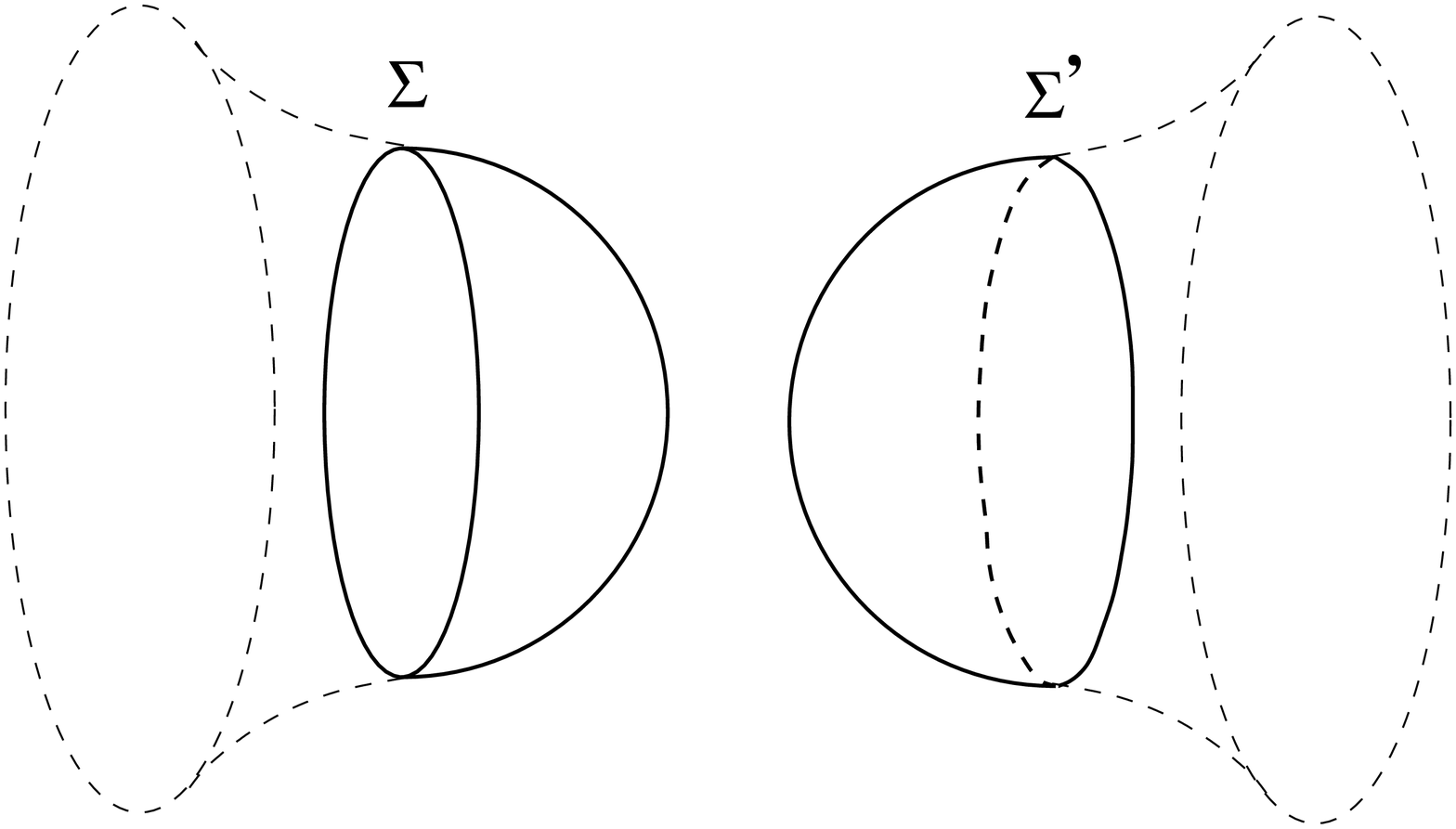}
\caption{Density matrix of the pure Hartle-Hawking state represented by the
union of two vacuum instantons.}
\end{figure}
The relevant density matrix is the path integral
\begin{equation}
\rho[\,\varphi,\varphi'\,]=\mbox{$e$}^{
    \Gamma}\!\!\!\!\!\!\!\!\!\!\!\!\!\!
    \int\limits_{\,\,\,\,\,\,\,\,g,\,
    \phi\,|_{\,\Sigma,\Sigma'}\,=\,(\,\varphi,\varphi')}
    \!\!\!\!\!\!\!\!\!D[\,g,\phi\,]\,
    \exp\big(-S_{\rm E}[\,g,\phi\,]\big).
\end{equation}
with the partition function $e^{-\Gamma}$ which follows from
integrating out the field $\varphi$ in the coincidence
$\varphi'=\varphi$ corresponding to the identification of $\Sigma'$
and $\Sigma$, the underlying instanton acquiring the toroidal
topology.

The metric of the instanton introduced above
is conformally equivalent to the metric of the Einstein static
universe:
\begin{equation}
d\bar{s}^2 = d\eta^2 + d^2\Omega^{(3)},
\label{Ein-stat1}
\end{equation}
where $\eta$ is the conformal time parameter. We shall consider
conformally invariant fields. As is well known, the quantum
effective action for such fields has a conformal anomaly first
studied in cosmology in \cite{Starob,Hu}. It has the form
\begin{equation}
g_{\mu\nu}\frac{\delta
    \Gamma_{\rm 1-loop}}{\delta g_{\mu\nu}} =
    \frac{1}{4(4\pi)^2}g^{1/2}
    \left(\alpha \Delta R +
    \beta E +
    \gamma C_{\mu\nu\alpha\beta}^2\right),        \label{anomaly}
\end{equation}
where $E = R_{\mu\nu\alpha\gamma}^2 -4R_{\mu\nu}^2 + R^2$ and $\Delta$
is the four-dimensional Laplacian. This anomaly, when integrated
functionally along the orbit of the conformal group, gives the
relation between the actions on conformally related backgrounds
\cite{BMZ}.
\begin{eqnarray}
    &&\Gamma_{\rm 1-loop}[\,g\,]=
    \Gamma_{\rm 1-loop}[\,\bar g\,]+\delta\Gamma[\,g,\bar
    g\,],\,\,\,
\\&&g_{\mu\nu}(x)=e^{\sigma(x)}\bar g_{\mu\nu}(x),
    \end{eqnarray}
where
\begin{eqnarray}
    &&\delta\Gamma[\,g,\bar g\,]
    =
    \frac{1}{2(4\pi)^2}\int d^4x \bar g^{1/2} \left\{\,\frac{1}{2}\,
    \Big[\,\gamma\, \bar C_{\mu\nu\alpha\beta}^2\right.\nonumber \\
        &&
    +\beta\,\Big(\bar E-\frac{2}{3}\,\bar{\Delta} \bar R\Big)\Big]\,
    \sigma                                  \nonumber\\
    &&
    \left.+\,\frac{\beta}{2}\,\Big[\,(\bar\Delta\sigma)^2
    +\frac{2}{3}\,\bar R\,(\bar\nabla_{\mu}\sigma)^2\,
    \Big]\,\right\}\nonumber\\
    &&-
    \,\frac{1}{2(4\pi)^2}\Big(\frac{\alpha}{12}
    +\frac{\beta}{18}\Big)\,\nonumber \\
    &&\times\int d^4x\,\Big(g^{1/2}R^2(g)-
    \bar{g}^{1/2}R^2(\bar{g})\Big).              \label{deltaW}
    \end{eqnarray}

One can show that the higher-derivative in $\sigma$ terms are all
proportional to the coefficient $\alpha$. The $\alpha$-term can be
arbitrarily changed by adding a local counterterm $\sim g^{1/2}R^2$.
We fix this local renormalization ambiguity by an additional
criterion of the absence of ghosts. The conformal contribution to
the renormalized action on the minisuperspace background equals
    \begin{eqnarray}
    &&\delta \Gamma[\,g,\bar g\,]\equiv
    \Gamma_{R}[\,g\,]
    -\Gamma_{R}[\,\bar g\,]\nonumber \\
&&=
    m_P^2\,B\!\int d\tau
    \left(\frac{\dot{a}^2}{a}
    -\frac16\,\frac{\dot{a}^4}a\right),   \label{correction}\\
    &&m_P^2\,B=\frac34\,\beta,       \label{Bm_P^2}
    \end{eqnarray}
with the constant $m_P^2\,B$ which for scalars, two-component
spinors and vectors equals respectively $1/240$, $11/480$ and
$31/120$.
For a conformal scalar field
\begin{equation}
    S[\,\bar g,\phi\,] = \frac12\,
    \sum_{\omega}\int_0^{\eta}
    d\eta' \left(\Big(\frac{d\phi_\omega}{d\eta'}\Big)^2+
    \omega^2\,\phi^2_\omega\right),                   \label{scal-action}
    \end{equation}
where $\omega=n$, $n=0,1,2,...$, labels a set of eigenmodes and
eigenvalues of the Laplacian on a unit 3-sphere. Thus
\begin{eqnarray}
    &&\mbox{$e$}^{\textstyle
    -\Gamma_{\rm 1-loop}[\,\bar g\,]}\nonumber \\&&=\int \prod_\omega
    d\varphi_\omega
    \!\!\!\!\!\!\!\!\!\!\!\!\!\!\!\!\!\!\!\!\!\!
    \int\limits_{\,\,\,\,\,\,\,\,\,\,\,\,\,\,\,\,
    \,\,\,\,\,\,\,\,\,\,\,\,
    \phi_\omega(\eta)=\phi_\omega(0)=\varphi_\omega}
    \!\!\!\!\!\!\!\!\!\!\!\!\!\!\!\!\!\!
    \!\!\!\!\!\!
    D[\,\phi\,]\,
    \exp\big(-S[\,\bar
    g,\phi\,]\big)
    \nonumber\\
    &&
    ={\rm const}\,\prod_\omega\left(\sinh
    \frac{\omega\eta}2\right)^{-1},
    \end{eqnarray}
and the effective action equals the sum of contributions of the
vacuum energy $E_0$ and free energy $F(\eta)$ with the inverse
temperature played by $\eta$ --- the circumference of the toroidal
instanton in units of a conformal time,
    \begin{eqnarray}
    &&\Gamma_{\rm 1-loop}[\,\bar g\,]
    =\sum_{\omega}\left[\,\eta\,
    \frac{\omega}{2}
    +\ln\big(1-e^{-\omega\eta}\big)\,\right]\nonumber \\&&=
    m_P^2\,E_0\,\eta+F(\eta),            \label{1000}\\
    &&m_P^2\,E_0=\sum_{\omega}
    \frac{\omega}{2}=\sum_{n=1}^\infty
    \frac{n^3}{2},                           \label{E_0}\\
    &&F(\eta)=\sum_{\omega}
    \ln\big(1-e^{-\omega\eta}\big)\\
    &&=\sum_{n=1}^\infty n^2\,
    \ln\big(1-e^{-n\eta}\big).
    \end{eqnarray}

Similar expressions hold for other conformally invariant fields of
higher spins. In particular, the vacuum energy (an analog of the
Casimir energy) on Einstein static spacetime is
\begin{eqnarray}
    m_P^2\,E_0=\frac1{960}\times\left\{\begin{array}{c} 4 \\
    17\\
    88\end{array}\right.
    \end{eqnarray}
respectively for scalar, spinor and vector fields.

We should take into account the effect of the finite ghost-avoidance
renormalization denoted below by a subscript $R$, which results in
the replacement of $E_0$ above by a new parameter $C_0$:
\begin{eqnarray}
    &&\Gamma_{R}[\,\bar g\,]
    =m_P^2\,C_0\,\eta_0+F(\eta),          \label{GammaRR}\\
    &&m_P^2\,C_0=m_P^2\,E_0+\frac3{16}\,\alpha.  \label{GammaR}
    \end{eqnarray}

A direct observation indicates the following universality relation
for all conformal fields of low spins
\begin{eqnarray}
    m_P^2\,C_0=\frac12\,m_P^2\,B.         \label{universality}
    \end{eqnarray}
Now we can write down the effective Friedmann equation governing the
Euclidean evolution of the universe. First of all, the full
conformal time on the instanton is
\begin{equation}
    \eta = 2\int_{\tau_-}^{\tau_+}
    \frac{d\tau\,N(\tau)}{a(\tau)},                \label{fulltime}
    \end{equation}
where $\tau_\pm$ label the turning points for $a(\tau)$ -- its
minimal and maximal values.

The effective action is ($m_P^2\equiv 3/4\pi G$)
\begin{eqnarray}
    &&\Gamma[\,a(\tau),N(\tau)\,]\nonumber \\&&=
    2 m_P^2\int_{\tau_-}^{\tau_+} d\tau\left(-\frac{a\dot{a}^2}N
    - Na + N H^2 a^3\right)\nonumber\\
    &&
    +2B m_P^2\int_{\tau_-}^{\tau_+}
    d\tau \left(\frac{\dot{a}^2}{Na}
    -\frac16\,\frac{\dot{a}^4}{N^3 a}\right)\nonumber\\
    &&
    +F\left(2\int_{\tau_-}^{\tau_+}
    \frac{d\tau\,N}{a}\right)+ B m_P^2
    \int_{\tau_-}^{\tau_+}
    \frac{d\tau\,N}{a}\,,
    \end{eqnarray}
and the effective Friedmann equation reads
\begin{eqnarray}
    &&\frac{\delta\Gamma}{\delta N}=
    2m_P^2\left(\frac{a\dot{a}^2}{N^2}
    - a + H^2 a^3\right)
    \nonumber\\
    &&
    +2Bm_P^2\left(-\frac{\dot{a}^2}{N^2 a}
    +\frac12\,\frac{\dot{a}^4}{N^4 a}\right)\nonumber \\&&
    +\frac2a \left(\frac{dF(\eta)}{d\eta}+\frac{B}2
    m_P^2\right)=0.
    \end{eqnarray}
In the gauge $N = 1$ this equation takes form
\begin{eqnarray}
    \frac{\dot{a}^2}{a^2}
    +B\,\left(\frac12\,\frac{\dot{a}^4}{a^4}
    -\frac{\dot{a}^2}{a^4}\right) =
    \frac{1}{a^2} - H^2 -\frac{C}{ a^4},     \label{Friedmann}
    \end{eqnarray}
where the amount of radiation constant $C$ is given by the bootstrap
equation
   \begin{equation}
    m_P^2 C = m_P^2\frac{B}2 +\frac{dF(\eta)}{d\eta}
    \equiv \frac{B}2 m_P^2+
    \sum_{\omega}\frac{\omega}{e^{\omega\eta}-1}. \label{bootstrap}
    \end{equation}

The Friedmann equation can be rewritten as
\begin{eqnarray}
    &&\dot{a}^2 = \sqrt{\frac{(a^2-B)^2}{B^2}
    +\frac{2H^2}{B}\,(a_+^2-a^2)(a^2-a_-^2)}\nonumber \\
    &&-\frac{(a^2-B)}{B}          \label{time-der1}
    \end{eqnarray}
and has the same two turning points $a_\pm$ as in the classical case
provided
    \begin{equation}
    a_-^2 \geq B.                \label{require}
    \end{equation}
This requirement is equivalent to
\begin{equation}
    C \geq B-B^2 H^2,\,\,\,\,B H^2\leq\frac12. \label{restriction1}
    \end{equation}
Together with
\[CH^2 \leq \frac{1}{4},\]
the admissible domain for instantons reduces to the curvilinear
wedge below the hyperbola and above the straight line to the left of
the critical point (see Figure 4)
\[C = \frac{B}{2},\ \ H^2 = \frac{1}{2B}.\]
\begin{figure}
\includegraphics[height=.4\textheight]{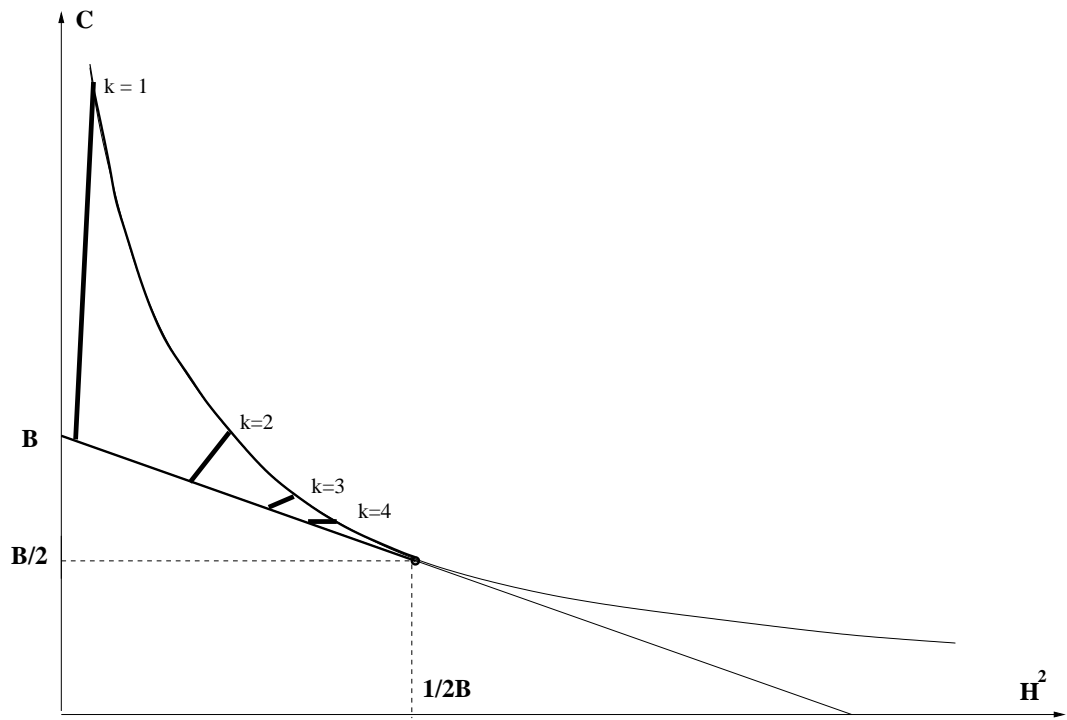}
\caption{The instanton domain in the $(H^2,C)$-plane is located
between bold segments of the upper hyperbolic boundary and lower
straight line boundary. The first one-parameter family of instantons
is labeled by $k=1$. Families of garlands are qualitatively shown
for $k=2,3,4$. $(1/2B,B/2)$ is the critical point of accumulation of
the infinite sequence of garland families.}
\end{figure}

The suggested approach allows to resolve the problem of the
so-called infrared catastrophe for the no-boundary state of the
Universe based on the Hartle-Hawking instanton. This problem is
related to the fact that the Euclidean action on this instanton is
negative and inverse proportional to the value of the effective
cosmological constant. This means that the probability of the
universe creation with an infinitely big size is infinitely high. We
shall show now that the conformal anomaly effect allows one to avoid
this counter-intuitive conclusion.

Indeed, outside of the admissible domain for the instantons with two
turning points, obtained above, one can also construct instantons
with one turning point which smoothly close at $a_- = 0$ with $\dot
a(\tau_-)=1$. Such instantons correspond to the Hartle-Hawking pure
quantum state. However, in this case the on-shell effective action,
which reads for the set of solutions obtained above as
    \begin{eqnarray}
    &&\Gamma_0= F(\eta)-\eta\frac{dF(\eta)}{d\eta}
    \nonumber \\
    &&
    +4m_P^2\int_{a_-}^{a_+}
    \frac{da \dot{a}}{a}\left(B-a^2
    -\frac{B\dot{a}^2}{3}\right),              \label{action-instanton}
    \end{eqnarray}
diverges to plus infinity. Indeed, for $a_-=0$ and $\dot a_{-}=1$
\begin{eqnarray}
&&\eta = \int_0^{a_+}\frac{da}{\dot a
a}=\infty,\,\,\,F(\infty)=F'(\infty)=0,
\end{eqnarray}
and hence the effective Euclidean action diverges at the lower limit
to $+\infty$. Thus,
\[\Gamma_0 = +\infty,\ \ \exp(-\Gamma_0) = 0,\]
and this fact completely rules out all pure-state instantons,
and only mixed quantum states of the universe, described by the cosmological
density matrix appear to be admissible.

In connection with all said above 
a natural question arises: where Euclidean quantum
gravity comes from? The answer can be formulated briefly as follows:
from the Lorentzian quantum gravity (LQG) \cite{Bar-dens}. Namely,
the density matrix of the Universe for the microcanonical ensemble
in Lorentzian quantum cosmology of spatially closed universes
describes an equipartition in the physical phase space of the
theory, but in terms of the observable spacetime geometry this
ensemble is peaked about a set of cosmological instantons (solutions
of the Euclidean quantum cosmology) limited to a bounded range of
the cosmological constant. These instantons obtained above as
fundamental in Euclidean quantum gravity framework, in fact, turn
out to be the saddle points of the LQG path integral, belonging to
the imaginary axis in the complex plane of the Lorentzian signature
lapse function \cite{Bar-dens}.

Now let us consider the cosmological evolution of the unverse
starting from the initial conditions described above. Making the
transition from the Euclidean time to the Lorentzian one, $\tau=it$,
we can write the modified Lorentzian Friedmann equation as
\cite{Bar-Kam-Def}
    \begin{eqnarray}
    &&\frac{\dot{a}^2}{a^2} + \frac{1}{a^2} =
    \frac{1}{B}\left\{1 - \sqrt{1-\frac{16\pi G}3\,B\,
    \varepsilon}\,\right\},\\
    &&\varepsilon=\frac3{8\pi G}\left(H^2+\frac{\cal
    C}{a^4}\right),\\
    &&{\cal C} \equiv C -\frac{B}{2}, \label{Lor-ev}
    \end{eqnarray}
where $\varepsilon$ is a total gravitating matter density in the
model (including at later stages also the contribution of particles
created during inflationary expansion and thermalized at the
inflation exit). A remarkable feature of this equation is that the
Casimir energy is totally screened here and only the thermal
radiation characterized by $\cal C$ weighs.

If one wants to compare the evolution described by Eq.
(\ref{Lor-ev}) with the real evoltuion of the universe, first of all
it is necessary to have a realistic value for an effective
cosmological constant $\Lambda=3H^2$. The only way to achieve this
goal is to increase the number of conformal fields and the
corresponding parameter $B$, (\ref{Bm_P^2}), of the conformal
anomaly (\ref{anomaly}). The mechanisms for growing number of the
conformal fields exist in some string inspired cosmological models
with extra dimensions \cite{Bar-dens}. If some of these mechanisms
work we can encounter an interesting phenomenon: if the $B$ grows
with $a$ faster than the rate of decrease of the energy density
$\varepsilon$ one encounters a new type of the cosmological
singularity - Big Boost. This singularity is characterized by finite
values of the cosmological radius $a_{BB}$ and of its time
derivative $\dot{a}_{BB}$, while the second time variable $\ddot{a}$
has an infinite positive value. The universe reaches this
singularity at some finite moment of cosmic time $t_{BB}$:
\begin{eqnarray}
&&a(t_{BB}) = a_{BB} < \infty,\\
&&\dot{a}(t_{BB}) = \dot{a}_{BB} < \infty,\\
&&\lim_{t \rightarrow t_{BB}} \ddot{a}(t) = \infty.
\label{BigBoost}
\end{eqnarray}

In paper \cite{Bar-Kam-Def1} it was found that there exist some correspondences between quantum 4-dimensional 
equations of motion and some classical 5-dimensional equations of motion \cite{Bar-Kam-Def,Bar-Kam-Def1}.There were   considered two five-dimansional models: the Randall -Sundrum model\cite{RS} and
the generalized Dvali-Gabadadze-Porrati (DGP) model \cite{DGP}. 

The Randall-Sundrum braneworld model
is a 4-dimensional spacetime braneworld embedded into the 5-dimensional anti-de Sitter 
bulk with the radius $L$. 
In the limit of small energy densities the modified quantum Friedmann equations coincide 
with the modified 4-dimensional Friedmann equations of the Randall-Sundrum model 
provided
\begin{equation}
\beta G = \frac{\pi L^2}{2}. 
\end{equation}

The 5-dimensional action of the generalized DGP model includes 
the 5-dimensional curvature term, the 5-dimensional cosmological constant 
and the 4-dimensinal curvature term on the brane. 

If we require the spherical symmetry, when we have the Schwarzschild-de Sitter solution,
which depends also on the Schwarzschil radius $R_S$. 
The effective 4-dimensional Friedmann equations on the 4-brane coincide with 
the modified Friedmann equations in quantum model, provided the quantity of the 
radiation is expressed through the Schwarzschild radius as 
\begin{equation}
C = R_S^2
\end{equation}
If we add the condition of the regularity of the Schwarzschild-de Sitter instanton,
(i.e. the condition of the absence of conical singularities),
we obtain an additional relation for the parameters 
of the quantum cosmological model
and the set of
admissible values for the effective cosmological
constant becomes discrete. 

Concluding this section, we would like to say that relaxing the usual tacit requirement 
of the purity of the quantum state of the universe and imposing the conditions of quantum consistency of the system of 
equations governing the dynamics of the universe, one comes to non-trivial restrictions on the basic cosmological paprameters.
Besides, as a by-product one obtains a particular kind of future soft singularity - Big Boost. 
Finally, we can note that in the papers, reviewed in this section both the main approaches 
to the study of quantum effects in cosmology were combined -- the study of the modified Friedmann equations and the investigation of the structure of the quantum state of the universe. Usually, these two approaches are separated (see, Sec. 11 and Secs. 9 and 10 of the present review). 

\section{Quiescent singularities in braneworld models}

One of the first examples of the soft future singularities in cosmology was presented in paper \cite{Shtanov},
where some braneworld cosmological models were considered. The higher-dimensional models considered there were described by an action, where both the bulk and brane contained the corresponding curvature terms:
\begin{equation}
\hspace{-1.5cm}S = M^3\sum_i\int_{\rm bulk}({\cal R}- 2\Lambda_i) - 2\int_{\rm brane} K +\int_{\rm brane}(m^2R-2\sigma)+\int_{\rm brane}L(h_{\alpha\beta},\phi),
\label{braneaction}
\end{equation}
where the sum is taken over the bulk components bounded by branes, and $\Lambda_i$ is the cosmological constant on the ith bulk component. The Lagrangian $L(h_{\alpha\beta},\phi)$ corresponds to the presence of matter fields on the brane interacting with the induced metric $h_{\alpha\beta}$, $K$ is the trace of the extrinsic curvature. 
The Friedmann-type equation has the form 
\begin{equation}
H^2 +\frac{\kappa}{a^2} = \frac{\rho+\sigma}{3m^2}+\frac{2}{l^2}\left[1\pm\sqrt{1+l^2\left(\frac{\rho+\sigma}{3m^2}-\frac{\Lambda}{6}-\frac{C}{a^4}\right)}\right],
\label{braneFried}
\end{equation}
where $\rho$ is the energy density of the matter on the brane, the integration constant $C$ corresponds to the presence of a black hole in the five-dimensional bulk solution, and the term $C/a^4$, sometimes called ``dark radiation'', arises due to the projection of the bulk gravitational degrees of freedom onto the brane. The length scale $l$ is defined as 
\begin{equation}
l = \frac{m^2}{M^3}.
\end{equation}
The appearence of the quiescent singularities is conneced with the fact, that the expression under the square root in (\ref{braneFried}) turns to zero at some point during the evolution. There are essentially two types of singularities dispaying this behaviour. 

A type 1 singularity (S1) is induced by the presence of the dark radiation term and arises in either of the following two cases:
$C > 0$ and the density of matter increases slower than $a^{-4}$ as $a \rightarrow 0$. An example is provided by dust. \\
The energy density of the universe is radiation dominated so that $\rho=\rho_0/a^4$ and $C > \rho_0$.

These singularities can take place either in the past of an expanding universe or in the future of a collapsing one. 

A type 2 singularity (S2) arises if 
\begin{equation}
l^2\left(\frac{\sigma}{3m^2}-\frac{\Lambda}{6}\right) < -1.
\label{brane2}
\end{equation}
In this case the combination $\rho/3m^2-C/a^4$ decreases monotonically as the universe expands. The expression under the square root of (\ref{braneFried}) can therefore become zero at suitably late times. 

For both S1 and S2, the scale factor $a(t)$ and its first time derivative remain finite, while all the higher time derivatives 
of $a$ tend to infinity as the singularity is approached. It is important that the energy density and the pressure of the matter in the bulk remain finite. This feature distinguishes these singularities from the singularities considered in the preceding sections, and justifies the special name ``quiescent'' \cite{Shtanov}. The point is that the existence of these singularities is connected not with special features of the matter on the brane, but with the particularity of the embedding of the brane into the bulk. 

In paper \cite{Shtanov1} the question of influence of the quantum effects on a braneworld encountering a quiescent singularity during expansion was studied.   
The matter considered in \cite{Shtanov1} was constituted from conformally invariant fields. Hence, the  particle production was absent and the only quantum effect was connected with the vacuum polarization. It was shown that this effect boils down to the modification of the effective energy density of the matter on the brane. Namely, the quantum correction to this energy density is given by 
\begin{equation}
\rho_{\rm quantum} = k_2H^4+k_3(2\ddot{H}H+6\dot{H}H^2-\dot{H}^2).
\label{branequant}
\end{equation}
The insertion of this correction to the energy density changes drastically the form of the brane Friedmann-type equation 
(\ref{braneFried}) -- the original algebraic equation becomes a differential equation. It implies  essential changes in the possible behaviour of the universe around singularities. First, the quiscent singularity changes its form and becomes much weaker, in fact, $H$ and $\dot{H}$ remain finite and only $\stackrel{\cdots}{H} \rightarrow \infty$. Second, vacuum polarization effects can also cause a spatially flat universe to turn around and collapse. 

At the conclusion of this section we would like to mention another type of cosmological singularities, arising in the brane-world context. These are the so called pressure singularities \cite{pressure-sing,pressure-sing1}. 
These singularities arise in the generalized Friedmann branes, which can be asymmetrically embedded into the bulk and can 
include pull-backs on the brane some non-standard field and geometric configurations, existing in the 5-dimensional bulk \cite{Gen-brane}. It appears that it is possible to reproduce in this frame work a Swiss cheese Einstein-Strauss model \cite{Strauss}. In this model there pieces of the Schwarzschild regions inserted into a Friedmann universes. At some conditions 
in the Friedmann regions of such branes the pressure of matter becomes infinite, while the cosmological radius and all 
its time derivatives remain finite. It was shown also \cite{pressure-sing1} that at some critical value of the assymetry in the embedding of the brane into the bulk, these singularities appear necessarily. It is interesting that these pressure singularities are in a way complementary to the quiescent singularities, discussed above, where the energy density and the pressure are always finite, while the time derivatives of the scale factor become divergent, beginning since the second or some higher-order derivative.

\section{Concluding remarks}
In this review we have considered a broad class of phenomena arising in cosmological models, possessing some exotic 
cosmological singularities, which differ from the traditional Big Bang and Big Crunch singularities. We have discussed the models, based on standard scalar fields, Born-Infeld-type fields and on perfect fluids, where soft future cosmological singularities exist and are transversable. The crossing of such singularities (or other geometrically peculiar surfaces in the spacetime) can imply such an interesting phenomenon as a transformation of matter properties, which is discussed in some detail here. Another interesting aspect of the study of both soft and ``hard'' (Big Bang or Big Crunch) cosmological singularities
is the existence of the correspondence between the phenomenon of quantum avoidance (or non-avoidance) of such singularities and the possibility of their crossing (or the absence of such a possibility) in classical cosmology.

Besides, the quantum cosmological approach, based on the study of the properties of solutions of the Wheeler-DeWitt equation, we have reviewed also some works based on the investigation of the modification of the Friedmann equation due to the quantum corrections and the influence of of these corrections on the structure and the very existence of soft cosmological singularities.

While the main part of this review deals with the standard Einstein general relativity in the presence of non-standard matter, the last section is devoted to the exotic singularities arising in the brane-world cosmological models, which are very close in their nature to the soft sudden singularities arising in the general relativity.

Generally, we are convinced that the study of exotic singularities in classical and quantum cosmology is a promising branch of the theoretical physics, and nobody can exclude that it can acquire some phenomenological value as well.
Here it is necessary to recognize that almost all studies in this field deal only with isotropic and homogeneous Friedmann universes. Thus, the extension of this studies to the anisotropic and inhomogeneous models represents a main challenge for people working in this field. Such an extension can bring some interesting surprises as it was with the study of the Big Bang - Big Crunch singularities, where the consideration of the anisotropic Bianchi models instead of Friedmann models, has given birth to the discovery of the oscillating approach to the singularity (Mixmaster Universe) \cite{BKL, Misner}.

\ack
I am grateful to A.M. Akhmeteli, A.A. Andrianov, A.O. Barvinsky, V.A. Belisnky, M. Bouhmadi-Lopez, F. Cannata, S. Cotsakis, M.P. Dabrowski, C. Deffayet, G. Esposito, L.A. Gergely, V. Gorini, D.I. Kazakov, Z. Keresztes, I.M. Khalatnikov, C. Kiefer, V.N. Lukash, S. Manti, P.V. Moniz, U. Moschella, V. Pasquier, D. Polarski, D. Regoli, V.A. Rubakov, B. Sandhofer, D.V. Shirkov, A.A. Starobinsky, O.V. Teryaev, A.V. Toporensky, V. Sahni, G. Venturi, A. Vilenkin and A.V. Yurov for fruitful discussions. 
This work was partially supported by the RFBR grant 11-02-00643.

\end{document}